\documentclass{article}
\pdfoutput=1
\usepackage{arxiv}

\usepackage{natbib}
\usepackage[utf8]{inputenc} % allow utf-8 input
\usepackage[T1]{fontenc}    % use 8-bit T1 fonts
\usepackage{url}            % simple URL typesetting
\usepackage{booktabs}       % professional-quality tables
\usepackage{amsfonts}       % blackboard math symbols
\usepackage[dvips]{graphicx}
\usepackage{caption}
\usepackage{amssymb}
\usepackage{amsmath}
\usepackage{subfig} 
\usepackage{color}
\usepackage{bm}
\usepackage{framed} % or, "mdframed"
\usepackage[framed]{ntheorem}
\usepackage{algorithm,algorithmic}
\usepackage{multirow}
\usepackage{epstopdf} %only rquired on windows

\newcommand{\E}{\mathbb{E}}
\newcommand{\Var}{\mathrm{Var}}

\newcommand{\MSE}{\mathrm{MSE}}
 
\DeclareMathOperator*{\argmax}{arg\,max} 
\newframedtheorem{frm-thm}{Theorem}

\title{Exploring Positive Noise in Estimation Theory}

  	\author{Kamiar Radnosrati \\  Department of Electrical Engineering\\
  		  Link\"oping University\\Link\"oping, Sweden\\ \texttt{kamiar.radnosrati@liu.se} 
  		 \And Gustaf Hendeby \\ Department of Electrical Engineering\\
  		 Link\"oping University\\Link\"oping, Sweden\\\texttt{gustaf.hendeby@liu.se}  \And Fredrik Gustafsson\\  Department of Electrical Engineering\\
  		 Link\"oping University\\Link\"oping, Sweden\\\texttt{fredrik.gustafsson@liu.se} }
\begin{document}
\maketitle

\begin{abstract}
Estimation of a deterministic quantity observed in non-Gaussian additive noise is explored via order statistics approach.  More specifically, we study the estimation problem when measurement noises either have positive supports or follow a mixture of normal and uniform distribution. This is a problem of great interest specially in cellular positioning systems where the wireless signal is prone to multiple sources of noises which generally have a positive support. Multiple noise distributions are investigated and, if possible, minimum variance unbiased (MVU) estimators are derived. In case of uniform, exponential and Rayleigh noise distributions, unbiased estimators without any knowledge of the hyper parameters of the noise distributions are also given. For each noise distribution, the proposed order statistic-based estimator's performance, in terms of  mean squared error, is compared to the best linear unbiased estimator (BLUE), as a function of sample size, in a simulation study.
%Estimation of the mean of a stochastic variable observed in noise with positive support is considered. It is well known from the literature that order statistics give one order of magnitude lower estimation variance compared to the best linear unbiased estimator (BLUE). We provide a systematic survey of some common distributions with positive support, and provide derivations and summaries of estimators based on order statistics, including the BLUE one for comparison.  In case of uniform, exponential and Rayleigh noise distributions, unbiased estimators without any knowledge of the hyper parameters of the noise distributions are also given. For each noise distribution, the proposed order statistic-based estimator's performance, in terms of  variance, is compared to the best linear unbiased estimator (BLUE), as a function of sample size, in a simulation study. The results show that, even for unknown hyper parameter scenarios, the proposed estimators have less variance compared to BLUE.
\end{abstract}

% keywords can be removed

%%%%%%%%%%%%%%%%%%%%%%%%%%%%%%%%%%%%
%---------------------------SECTION----------------------------
%%%%%%%%%%%%%%%%%%%%%%%%%%%%%%%%%%%%
\section{Introduction}\label{sec:introduction}
We consider the problem of estimating the mean $x$ observed in noise as $y_k=x+e_k$, for $k=1,2,\dots, N$,  also known as ``estimation of location"~\citep{article:PI_kassam_85}, where the noise $e_k$ has positive support. We will refer to such distributions as {\em positive noise}. Examples of distributions we will study include uniform, exponential, Rayleigh, Pareto.

A bias compensated linear estimator as the sample mean has a variance that decays as $1/N$, while it is well-known from the statistical literature, see for example \citep{book:ET_kay_93,book:TPE_lehmann}, that the minimum has a variance that decays as $1/N^2$. The minimum is the simplest example of order statistics. Certain care has to be taken for the cases where the parameters in the distributions are unknown, in which case bias compensation becomes tricky. This paper derives all combinations of known/unknown parameters for order statistics/BLUE (best linear unbiased estimator) for some selected and common distribution that allow for analytical solutions. 

\begin{figure}[]
	\centering
	\subfloat{\includegraphics[width=0.33\columnwidth]{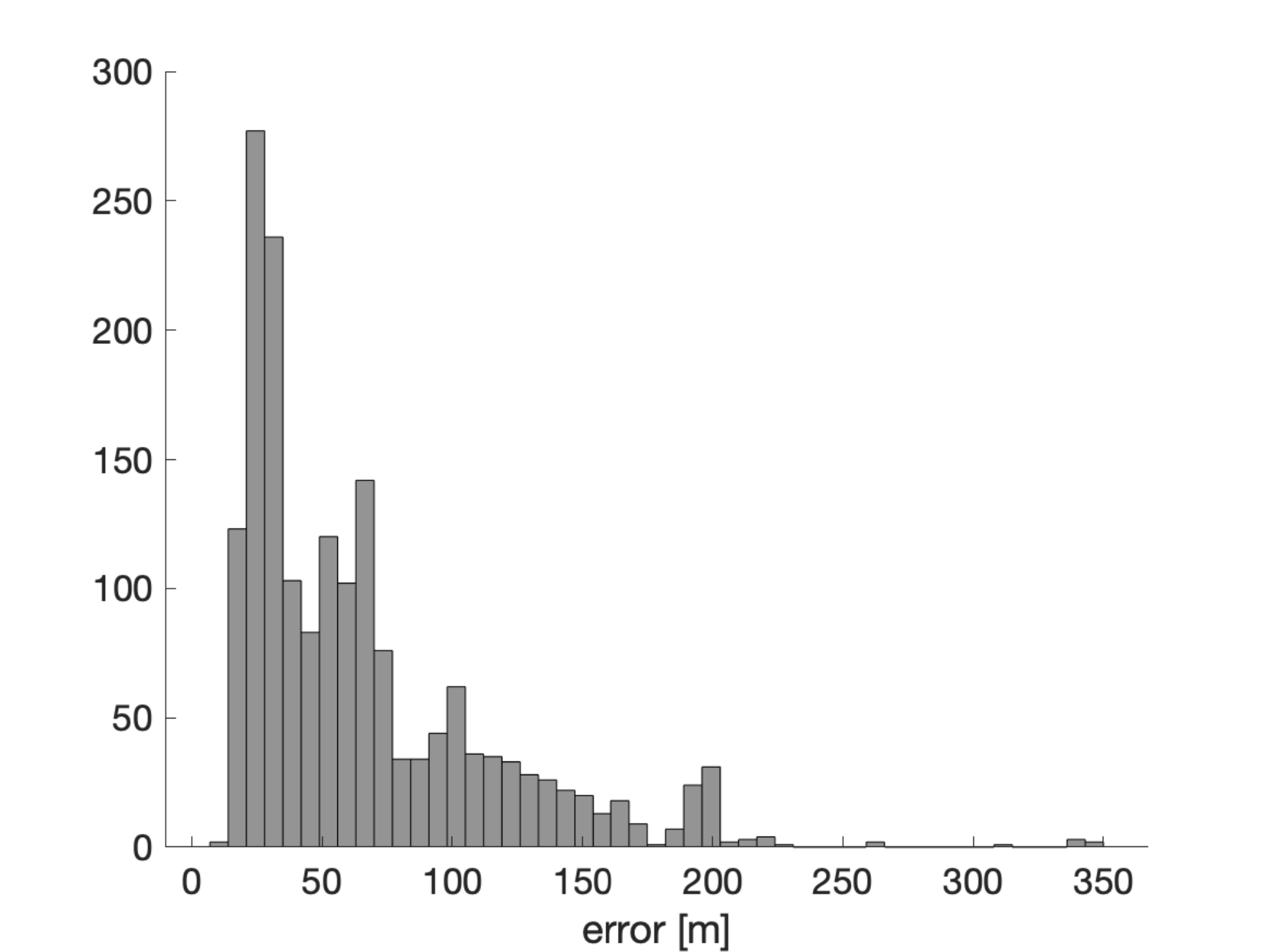}%
		\label{fig:kista_1}}
	\hfil
	\subfloat{\includegraphics[width=0.33\columnwidth]{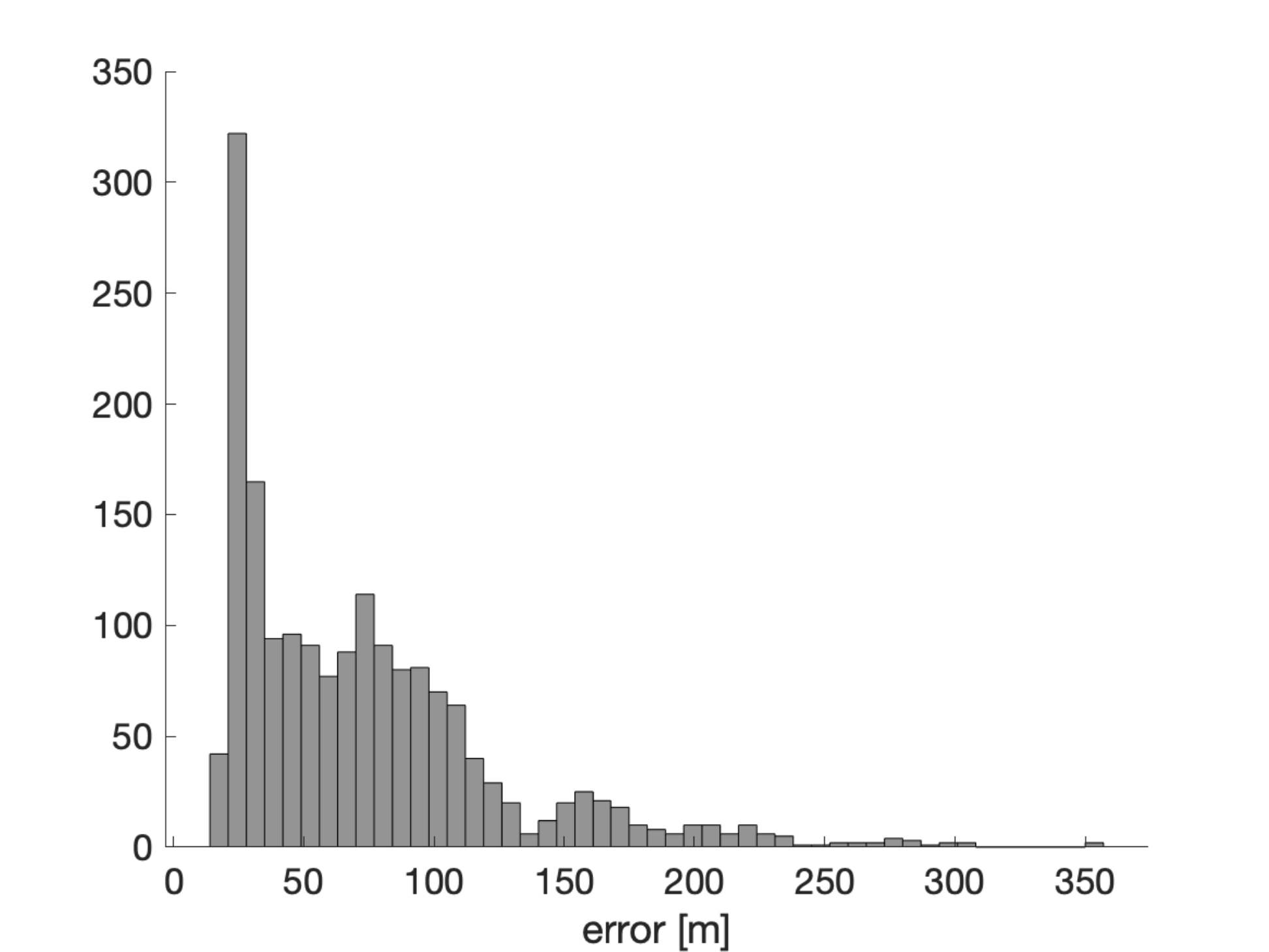}%
		\label{fig:kista_2}}
	\hfil
	\subfloat{\includegraphics[width=0.33\columnwidth]{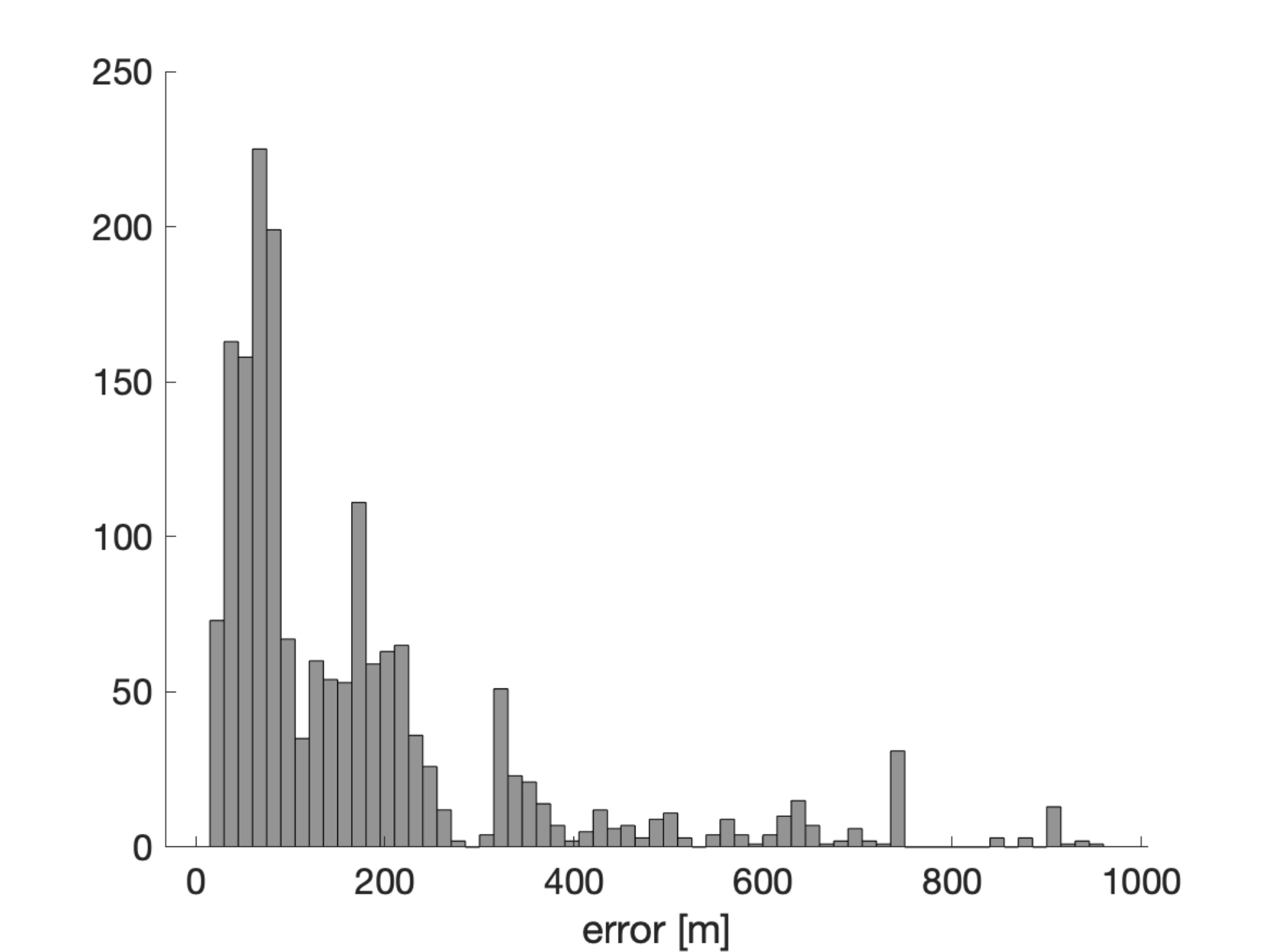}%
	\label{fig:kista_3}}	
	\caption{The error histograms of time-of-arrival measurements collected from three separate cellular antennas described in~\citep{conf:PIMRC_medbo_09}.}
	\label{fig:kista}
\end{figure}

Problems involving positive noise can be motivated from applications where the arrival times of radio or sound waves are used. Such waves travel with the speed of the medium, and non line of sight conditions give rise to delayed arrival times. Physics does simply not allow for negative noise, only positive one. This case occur in a variety of applications such as target tracking using radar or lidar, and localisation using radio waves such as is done in for instance global satellite navigation systems~\citep{article:ITVT_kok_15,article:ITVT_chen_09,article:ISPM_gustafsson_05,article:IME_eling_12}. For example, the error histograms of time-of-arrival measurements collected from three separate cellular antennas are given in Figure~\ref{fig:kista}. For detailed description of hardware and the measurement campaign see~\citep{conf:PIMRC_medbo_09}

To deal with the estimation's performance degradation in non-Gaussian error conditions, conventional estimation techniques which are developed based on Gaussian assumptions need to be adjusted properly. As discussed in~\citep{article:ITSP_yin_13}, ``identify and discard'', ``mathematical programming'', and ``robust estimation'' are the three broad categories of estimation methods which are robust against non-Gaussian errors. Robustness of the estimator has been a concern for many years in both research~\citep{article:JASA_stigler_73} and different engineering topics ~\citep{article:PI_kassam_85,book:SDNGN_kassam,article:SIAM_stewart_99,book:NSP_arce} for a long time now. A more recent survey on this topic containing more references can be found in~\citep{article:ISPM_zoubir_12}.

The maximum likelihood estimator (MLE), developed under Gaussian assumptions, can be modified to become robust in presence of non-Gaussian noises. The authors in~\citep{conf:ICML_eskin_00} first detect and then reject the outliers by learning the probability density function (PDF) of the measurements and develop a mixture model of outliers and clean data. A similar idea to k-nearest-neighbor approach is used in~\citep{article:DMKD_chawla_10} to classify outliers as the data points that are less similar to a number of neighboring measurements. Surveys of advances in clustering the data into outliers and clean data can be found in~\citep{article:AIR_hodge_04,article:ITSP_yin_13,conf:ICASSP_fritsche_09}. While these approaches might result in high estimation accuracy, they typically require large datasets~\citep{article:ISPM_zoubir_12}.

M-estimators~\citep{book:RS_huber}, in the contrary to identification-based methods, do not require pre-processing and can be used in non-Gaussian noise conditions. In principle, M-estimators can be seen as generalization of MLE and rely on solving a minimization problem of some loss function. For a detailed discussion on different loss functions, see~\citep{book:RS_huber}. Since minimization problems are typically solved numerically based on the derivative of the loss function~\citep{book:RSTM_maronna}, they might converge to local minima.

In this work, we strive to find minimum variance unbiased (MVU) estimators for the location of estimation problems for non-Gaussian noise distributions where multiple distributions with positive support are considered.  In case where MVU is not found, we introduce unbiased order-statistic-based estimators and compare their variances against the BLUE. The MVU estimators without any knowledge of the hyper parameters of the noise distributions are also derived, if possible. Finally, we derive an estimator for the case in which the noise follows a mixture of normal and uniform distribution.
%
%The most intuitive way for finding the MVU is to use Cramer-Rao lower bound (CRLB) theorem. However, the considered noise distributions do not satisfy regularity condition, hence CRLB theorem cannot be applied. Instead, we rely on  Rao-Blackwell-Lehmann-Scheffe (RBLS) theorem to find the MVU. Since the PDF cannot always be factorized in the framework of the RBLS theorem, however, we are not able to find MVU in all scenarios. 

The rest of this paper is structured  as follows. In Section~\ref{sec:marginal_distribution_of_order_statistics} the marginal distribution of order statistics is introduced. In Section~\ref{sec:location_estimation_problem} the location estimation problem is formulated. The problem is then investigated for different noise distributions and estimators for each distribution are derived in Sections~\ref{sec:uniform_distribution}--\ref{sec:other_distributions}. The proposed estimators are evaluated in a simulation study in Section~\ref{sec:performance_evaluation} followed by the concluding remarks given in Section~\ref{sec:conclusions}.

%%%%%%%%%%%%%%%%%%%%%%%%%%%%%%%%%%%%
%---------------------------SECTION----------------------------
%%%%%%%%%%%%%%%%%%%%%%%%%%%%%%%%%%%%
\section{Marginal Distribution of Order Statistics}\label{sec:marginal_distribution_of_order_statistics}
The marginal distribution of order statistics, in this work is computed by differentiating the corresponding cumulative distribution function (CDF). In this section, we first introduce the minimum, also know as first or extreme, order statistic and then give the generalization to any statistics of order $k$. Let $F$ denote the common CDF of $N$ independent and identically distributed sample of random variables $y_1,\ldots,y_N$. We let $y_{(k)}$ denote the $k$:th order statistic of the sample, defined as the $k$:th smallest value of the set $\{y_i\}_{i=1}^N$. We define $f_{(k,N)}(y)$ as the marginal PDF of the $k$:th order statistics corresponding to a sample of size $N$. The PDF $f_{(k,N)}(y)$ is then calculated by differentiating $F_{(k,N)}(y)$ with respect to $y$.

%<<<<<<<<<<<<<<<<<<<<<<<<<<<<<<<<<<<<<<<<<<<<<<<<
%------------------------SUBSECTION--------------------------
%<<<<<<<<<<<<<<<<<<<<<<<<<<<<<<<<<<<<<<<<<<<<<<<<
\subsection{Marginal distribution of minimum order statistic}\label{subsec:marginal_distribution}
To further illustrate the problem, consider fist an example in which we have drawn $N=5$ independent random variables $\{y_i\}_{i=1}^5$ each from a common distribution with PDF $f(y)$.  Assume that we are interested in the PDF of the first order statistic, $f_{(1,5)}(y)$. The CDF $F_{(1,5)}(y)$ is defined as $P(y_{(1)} < y)$. We note that the minimum order statistic $Y_{(1)}$ would be less than $y$ if at least $1$ of the random variables $y_1, y_2, y_3, y_4, y_5$ are less than $y$. In other words, we need to count the number of ways that can happen such that at least one random variable is less than $y$. This leads to a binomial probability calculation. The 'success' is considered to be the event $\{y_i < y\}$, $i = 1$ and we let $\zeta$ denote the number of successes in five trials, then
\begin{align*}
F_{(1,5)}(y) &= P(y_{(1)}<y) = P(\zeta=1)+\ldots + P(\zeta=5),\\
f_{(1,5)}(y) &= \frac{\,d}{\,dy}F_{(1,5)}(y).
\end{align*}

To generalize the example, let $y_{(1)} < y_{(2)} < \ldots < y_{(N)}$ be the order statistics of $N$ independent observations from a continuous distribution with cumulative distribution function $F(y)$ and probability density function $f(y)=F'(y)$. The marginal PDF $f_{(1,N)}(y)$ of the minimum order statistic  can be obtained by considering the event $\{Y_i \leq y\}, i = 1$ as a "success," and letting $\zeta$ = the number of such successes in $N$ mutually independent trials. $\zeta$ is a binomial random variable with $N$ trials and probability of success $P(y_i\leq y)$. Hence, the CDF of the minimum order statistic is given by,
\begin{subequations}
	\begin{align}
	F_{(1,N)}(y)=\sum_{n=1}^{N}P(\zeta=n).
	\label{eq:cdf_order_minimum}
	\end{align}
	Noting that the probability mass function of this binomial distribution is given by,
	\begin{align}
	P(\zeta=n) = \begin{pmatrix}N\\n\end{pmatrix}[F(y)]^n[1-F(y)]^{N-n}.
	\label{eq:pmf_order_minimum}
	\end{align}

	Substituting~\eqref{eq:pmf_order_minimum} into~\eqref{eq:cdf_order_minimum} and taking the last term out of the sum, we get 
	\begin{align}
	F_{(1,N)}(y)=\sum_{n=1}^{N-1}\begin{pmatrix}N\\n\end{pmatrix}[F(y)]^n[1-F(y)]^{N-n}+[F(y)]^N.
	\label{eq:cdf_order_minimum_2}
	\end{align}
\end{subequations}

Differentiating~\eqref{eq:cdf_order_minimum_2} with respect to $y$ gives a telescoping sum of the form,
\begin{align}
f_{(1,N)}(y)&=\sum_{n=1}^{N-1}\frac{N!}{(n-1)!(N-n)!}[F(y)]^{n-1}f(y)[1-F(y)]^{N-n}\nonumber\\ &+ \sum_{n=1}^{N-1}\frac{N!}{n!(N-n-1)!}[F(y)]^n[1-F(y)]^{N-n-1}(-f(y)) \nonumber\\&+ N[F(y)]^{N-1}f(y),
\end{align}
in which, except the first term, all other terms cancel each other out. Hence, the marginal probability density function of the minimum order statistic of a set of $N$ independent and identically random variables with common CDF $F(y)$ and PDF $f(y)$ is given by,
\begin{align}
f_{(1,N)}(y) = Nf(y)\left(1-F(y)\right)^{N-1}.
\label{eq:density_order_minimum}
\end{align}
%
%
%<<<<<<<<<<<<<<<<<<<<<<<<<<<<<<<<<<<<<<<<<<<<<<<<
%------------------------SUBSECTION--------------------------
%<<<<<<<<<<<<<<<<<<<<<<<<<<<<<<<<<<<<<<<<<<<<<<<<
\subsection{Marginal distribution of general order statistic}\label{subsec:marginal_distribution_of_general_order_statistic}
The marginal PDF $f_{(k,N)}(y)$ of the general order statistic $k$ can be obtained by generalizing the results of the previous section, and considering the event $\{y_i \leq y\}, i = 1, 2, \ldots, k$ as a "success," and letting $\zeta$ = the number of such successes in $N$ mutually independent trials,
	\begin{align}
	F_{(k,N)}(y)=\sum_{n=k}^{N-1}\begin{pmatrix}N\\n\end{pmatrix}[F(y)]^n[1-F(y)]^{N-n}+[F(y)]^N.
	\label{eq:cdf_generic2}
	\end{align}
Differentiating~\eqref{eq:cdf_generic2} with respect to $y$ gives a telescoping sum of the form,
\begin{align}
f_{(k,N)}(y)&=\sum_{n=k}^{N-1}\frac{N!}{(n-1)!(N-n)!}[F(y)]^{n-1}f(y)[1-F(y)]^{N-n}\nonumber\\ &+ \sum_{n=k}^{N-1}\frac{N!}{n!(N-n-1)!}[F(y)]^n[1-F(y)]^{N-n-1}(-f(y)) \nonumber\\&+ N[F(y)]^{N-1}f(y),
\end{align}
in which, except the first term, all other terms cancel each other. Hence, the marginal probability density function of the $k$:th order statistic of a set of $N$ independent and identically random variables with common CDF $F(y)$ and PDF $f(y)$ is given by,
\begin{align}
f_{(k,N)}(y) = Nf(y)\begin{pmatrix}N-1\\k-1\end{pmatrix}F(y)^{k-1}\left(1-F(y)\right)^{N-k}.
\label{eq:density_order}
\end{align}
%---------------------------SECTION----------------------------
%%%%%%%%%%%%%%%%%%%%%%%%%%%%%%%%%%%%
\section{Location Estimation Problem}\label{sec:location_estimation_problem}
Consider the location estimation problem in which we have measurements $y_k$, $k=1,\ldots,N$ of the unknown parameter $x$. Assuming that the measurements are corrupted with additive noise $e_k\sim p_e(\theta)$, where $\theta$ denotes the parameter(s) of the noise distribution, the measurement model is given by
\begin{align}
y_k=x+e_k, \quad k=1,\ldots,N.
\label{eq:generic_model}
\end{align}
The BLUE for the estimation problem~\eqref{eq:generic_model} is given by
\begin{align}
\hat{x}_{\mathrm{BLUE}}^{p_e(\theta)}(y_{1:N},\theta) &= \frac{1}{N}\sum_{k=1}^N y_k- \delta(\theta),
\label{eq:mean_estimator}
\end{align}
where $y_{1:N} = \{y_k\}_{k=1}^{N}$ and $\delta(\theta)=\mathbb{E}(e_k)$ is the bias compensation term.

In the following sections, closed-form expressions for the mean squared error (MSE) of the BLUE estimator for multiple noise distributions with positive support are provided.  Given  hyperparameter $\theta$, the MVU estimator for each noise distribution $p_e$ is denoted by $\hat{x}_{\mathrm{MVU}}^{p_e}(y_{1:N},\theta)$. MVU estimators with unknown hyperparameter are denoted by  $\hat{x}_{\mathrm{MVU}}^{p_e}(y_{1:N})$. If the MVU cannot be found, an unbiased order-statistics-based estimator is derived and denoted by $\hat{x}^{p_e}(y_{1:N},\theta)$ and $\hat{x}^{p_e}(y_{1:N})$ for known and unknown hyperparameter cases, respectively. For example, $\hat{x}^{\mathcal{U}}_{\mathrm{MVU}}(y_{1:N},\beta)$ denotes the MVU estimator when $e_k\sim\mathcal{U}[0,\beta]$ and $\beta$  is known. $\hat{x}^{\mathcal{U}}_{\mathrm{MVU}}(y_{1:N})$, on the other hand, corresponds to the MVU estimator of uniform noise with unknown hyper parameters of the distribution. Table~\ref{tbl:notation} summarizes the notation used throughout this work.
\begin{table}[]
	\centering
	\caption{Notation.}
	\begin{tabular}{l|l}
		$\{y_k\}_{k=1}^N$                                                & noisy measurements of the unknown parameter $x$                                                          \\[2mm]
		$\left(y_{(m)}\right)_{m=1}^N$                                            & ordered measurement sequence                                          \\[2mm]
		$\theta$                                        & parameters of the noise distribution                                                                     \\[2mm]
		$\delta(\theta)$                                & bias compensation term                                                                                   \\[2mm]
		$\hat{x}^{p_e}_{\mathrm{BLUE}}(y_{1:N},\theta)$ & BLUE when $e_k\sim p_e$ for known $\theta$                                            \\[2mm]
		$\hat{x}^{p_e}_{\mathrm{MVU}}(y_{1:N},\theta)$  & MVU estimator when $e_k\sim p_e$ for known $\theta$                                   \\[2mm]
		$\hat{x}^{p_e}_{\mathrm{MVU}}(y_{1:N})$              & MVU estimator when $e_k\sim p_e$ for unknown $\theta$ \\[2mm]
		$\hat{x}^{p_e}(y_{1:N},\theta)$                 & unbiased estimator when $e_k\sim p_e$ for known $\theta$                              \\[2mm]
		$\hat{x}^{p_e}(y_{1:N})$                             & unbiased estimator when $e_k\sim p_e$ for unknown $\theta$                           
	\end{tabular}
	\label{tbl:notation}
\end{table}

For each noise distribution, we also consider the minimum order statistic estimator, denoted by $\hat{x}^{p_e}_{\mathrm{min}}(y_{1:N})$. Let $\left(y_{(m)}\right)_{m=1}^{N}$ denote the ordered sequence obtained from sorting $y_{1:N}$ in an ascending order, $\hat{x}^{p_e}_{\mathrm{min}}(y_{1:N})$ is defined as
\begin{align}
	\hat{x}^{p_e}_{\mathrm{min}}(y_{1:N}) = y_{(1)} \triangleq \min_{k} y_k.
\end{align}
Noting that for any generic estimator $\hat{x}$ the MSE is given by
\begin{align}\label{eq:mse_generic}
\MSE(\hat{x}) = \Var(\hat{x}) + b^2(\hat{x}).
\end{align}
The MSE for BLUE and MVU or any other bias compensated estimator coincides with the estimator's variance. In case of $\hat{x}^{p_e}_{\mathrm{min}}$, the existing bias enters the MSE.

In order to find the MVU estimator, the first step is to find the PDF $f(y_{1:N};\theta)$ with $\theta$ denoting the parameters of the distribution. If the PDF satisfies regularity conditions, the CRLB can be determined. Any unbiased estimator that satisfies CRLB is thus the MVU estimator. However, the considered PDFs do not satisfy the regularity conditions,
\begin{align}
\E\left[\frac{\partial\ln f(y_k;\theta)}{\partial \theta}\right]\neq0.
\end{align}
Hence, the CRLB approach is not applicable. Instead, we rely on the RBLS theorem~\citep{article:IJS_lehmann_1,article:IJS_lehmann_2,book:ET_kay_93}, to find the MVU estimator. The RBLS theorem~\citep{article:IJS_lehmann_1} states that for any unbiased estimator $\tilde{x}$ and sufficient statistics $T(y_{1:N})$, $\hat{x}=\E(\tilde{x}\mid T(y_{1:N}))$ is unbiased and $\Var(\hat{x})\leq\Var(\tilde{x})$. Additionally, if $T(y_{1:N})$ is complete, then $\hat{x}$ is MVU.

As shown in~\citep{book:ET_kay_93}, if the dimension of the sufficient statistics is equal to the dimension of the parameter, then the MVU estimator is given by $\hat{x}=g(T(y_{1:N}))$ for any function $g(\cdot)$ that satisfies
\begin{align}
\E(g(T)) = x.
\end{align}
Hence, the problem of MVU estimator turns into the problem of finding a complete sufficient statistic. The Neyman-Fisher theorem~\citep{article:fisher_22,article:AMS_halmos_49} gives the sufficient statistic $T(y_{1:N})$, if the PDF can be factorized as follows
\begin{align}
f(y_{1:N};\Psi) =g(T(y_{1:N}),\Psi)h(y_{1:N}),
\end{align}
where $\psi$ is the union of the noise hyper parameter(s) $\theta$ and $x$. The estimators in this work are derived in the order statistics framework. 
%
%
%%%%%%%%%%%%%%%%%%%%%%%%%%%%%%%%%%%%
%---------------------------SECTION----------------------------
%%%%%%%%%%%%%%%%%%%%%%%%%%%%%%%%%%%%
\section{Uniform Distribution}\label{sec:uniform_distribution}
As the first scenario, consider the case in which the additive noise $e_k$ has a uniform distribution with a positive support, $p_e(\theta)=\mathcal{U}[0,\beta]$, $\beta>0$ and $\theta=\beta$. The BLUE is given by
\begin{subequations}
	\begin{align}
	\hat{x}_{\mathrm{BLUE}}^{\mathcal{U}}(y_{1:N},\beta) &= \frac{1}{N}\sum_{k=1}^N y_k- \frac{\beta}{2}.
	\label{eq:sample_mean_example}
	\end{align}
	The MSE of BLUE for this case is given by
	\begin{align}
	\MSE\left(\hat{x}_{\mathrm{BLUE}}^{\mathcal{U}}(y_{1:N},\beta)\right) &= \frac{1}{N^2}\sum_{k=1}^{N}\Var\left(y_k-\frac{\beta}{2}\right)= \frac{\beta^2}{12N}.
	\label{eq:mse_blue_uniform}
	\end{align}
\end{subequations}

In order to find the MSE of the minimum order statistics estimator, $\hat{x}_{\mathrm{min}}^{\mathcal{U}}(y_{1:N}) $, we need to find the first two moments of the estimator.  Let $\tilde{y}_k= \frac{1}{\beta}y_k$. Since $y_k\sim\mathcal{U}[x,x+b]$, then for any constant $\beta>0$, $\tilde{y_k}\sim\mathcal{U}[\frac{x}{\beta},\frac{x}{\beta}+1]$. Hence, $f(\tilde{y}_k)=1$ and $F(\tilde{y}_k)=\frac{1}{\beta}(y_k-x)$. From~\eqref{eq:density_order} we get,
\begin{subequations}
	\begin{align}
	f&_{(k,N)}^{\mathcal{U}[0,\beta]}(\tilde{y}) = N\begin{pmatrix}N-1\\k-1\end{pmatrix}\left(\frac{\tilde{y}-x}{\beta}\right)^{k-1}\left(\frac{\beta-(\tilde{y}-x)}{\beta}\right)^{N-k}\nonumber\\
	&= \frac{(N)!}{(k-1)!(N-k)!}\left(\frac{\tilde{y}-x}{\beta}\right)^{k-1}\left(\frac{\beta-(\tilde{y}-x)}{\beta}\right)^{N-k}.
	\end{align}
	since $N\in\mathbb{N}^+$, $k\in\mathbb{N}^+$, and $k\in[1,N]$ we can the change the factorials to gamma functions,
	\begin{align}
	f_{(k,N)}^{\mathcal{U}[0,\beta]}(\tilde{y}) = \frac{\Gamma(N+1)}{\Gamma(k)\Gamma(N-k+1)} \left(\frac{\tilde{y}-x}{\beta}\right)^{k-1}\left(\frac{\beta-(\tilde{y}-x)}{\beta}\right)^{N-k}.
	\label{eq:order_uniform_general}
	\end{align}
\end{subequations}
The marginal distribution~\eqref{eq:order_uniform_general} is a generalized beta distribution, also known as four parameters beta distribution~\citep{article:Mcdonal_JE_95}. The support of this distribution is from $0$ to $\beta>0$ and $f_{(k,N)}^{\mathcal{U}[0,\beta]}(\cdot)=\frac{1}{\beta}f_{(k,N)}^{\mathcal{U}[0,1]}(\cdot)$.  The bias and variance of the general $k$:th order statistic estimator  $\hat{x}_{(k)}^{\mathcal{U}}(y_{1:N})$ in case of uniform noise with support on $[0,\beta]$ are given by
\begin{subequations}\label{eq:moments_uniform_order}
	\begin{align}
	\E(\hat{x}_{(k)}^{\mathcal{U}}(y_{1:N})) &= \frac{\beta k}{N+1},\\
	\Var(\hat{x}_{(k)}^{\mathcal{U}}(y_{1:N})) &= \frac{k(N-k+1)\beta^2}{(N+1)^2(N+2)}.
	\end{align}	
\end{subequations}
The first two moments of the minimum order statistic estimator are obtained by letting $k=1$ in~\eqref{eq:moments_uniform_order}
\begin{subequations}\label{eq:moments_uniform_min}
	\begin{align}
	b\left(\hat{x}_{\mathrm{min}}^{\mathcal{U}}(y_{1:N})\right) &= \frac{\beta}{N+1}\\
	\Var\left(\hat{x}_{\mathrm{min}}^{\mathcal{U}}(y_{1:N})\right) &= \frac{N\beta^2}{(N+1)^2(N+2)}.
	\end{align}
\end{subequations}
The MSE of $\hat{x}_{\mathrm{min}}^{\mathcal{U}}(y_{1:N})$ is then given by
\begin{align}
\MSE\left(\hat{x}_{\mathrm{min}}^{\mathcal{U}}(y_{1:N})\right) = \frac{2\beta^2}{(N+1)(N+2)}.
\label{eq:mse_min_uniform}
\end{align}
\subsection{MVU estimator}\label{subsec:mvu_estimator}
In order to find the MVU estimator, we note that the PDF can be written in a compact form using the step function $\sigma(\cdot)$ as
\begin{subequations}
	\begin{align}
	f(y_k;x,\beta) = \frac{1}{\beta}\left[\sigma(y_k-x) - \sigma(y_k-x-\beta)\right].
	\label{eq:uniform_pdf_single}
	\end{align}
	which gives
	\begin{align}
	f(y_{1:N};&x,\beta) = \frac{1}{\beta^N}\prod_{k=1}^{N}\left[\sigma(y_k-x) - \sigma(y_k-x-\beta)\right]\nonumber\\
	&=\frac{1}{\beta^N}\left[\sigma( y_{(1)} -x) - \sigma(y_{(N)}-x -\beta)\right],%\nonumber\\
	%&=g\left(T_1(\bm{y}),T_2(\bm{y}),x\right)\cdot h(\bm{y}),
	\label{eq:unknown}
	\end{align}
\end{subequations}
where $y_{(N)}\triangleq\max_k y_k,\quad k=1,\ldots,N$.
The expressions for the MVU estimator is derived for two different scenarios. We first assume that the hyper parameter $\beta$ of the noise distribution is known and then further discuss the unknown hyper parameter case.  In the general case, let $\Psi = [x,\beta]^\top$ denote the unknown parameter vector, the Neyman-Fisher factorization gives $h(y_{1:N})=1$ and 
\begin{align}
T(y_{1:N}) = \begin{bmatrix}
y_{(1)} \\ y_{(N)}
\end{bmatrix} = \begin{bmatrix}
T_1(y_{1:N}) \\ T_2(y_{1:N})
\end{bmatrix}.
\label{eq:uniform_ss}
\end{align}

\subsubsection{Known hyper parameter $\beta$}\label{subsubsec:known_hyper_parameter_uniform}
When the maximum support of the uniform noise $\beta$ is known, the dimensionality of the sufficient statistic is larger than that of the unknown parameter $x$. As discussed in~\citep{book:ET_kay_93}, the RBLS theorem can be extended to address this case if the form of a function $g(T_1(y_{1:N}),T_2(y_{1:N}))$ can be found that combines $T_1$ and $T_2$ into a single unbiased estimator of $x$. 

Let $Z = T_1(y_{1:N})+T_2(y_{1:N})=u+v$. Since $T_1$ and $T_2$ are dependent, 
\begin{subequations}
	\begin{align}\label{eq:dist_sum_gen}
	f_Z(z) = \int_{-\infty}^{\infty} f_{y_{(1)},y_{(N)}}(u,z-u)\,d_u,
	\end{align}
	where $f_{y_{(1)},y_{(N)}}(u,z-u)$ is the joint density of minimum and maximum order statistics. As shown in~\citep{book:OS_david_04}, for $-\infty<u<v<\infty$, the joint density of two order statistics $y_{(i)}$ and $y_{(j)}$ is given by
	\begin{align}
	f_{y_{(i)},y_{(j)}}(u,v) = &\frac{N!}{(i-1)!(j-1-i)!(N-j)!}\nonumber\\&\times f_Y(u)f_Y(v)\left[F_Y(u)\right]^{i-1}\nonumber\\
	&\times\left[F_Y(v)-F_Y(u)\right]^{j-1-i}\left[1-F_Y(v)\right]^{N-j},
	\end{align}
	that for the extreme orders, $i=1$ and $j=N$ can be simplified such that for $u<v$
	\begin{align}\label{eq:order_joint_gen}
	f_{y_{(1)},y_{(N)}}(u,v) = N(N-1)\left[F_Y(v)-F_Y(u)\right]^{N-2} f_Y(u)f_Y(v).	
	\end{align}
	and zero otherwise. Substituting~\eqref{eq:order_joint_gen} into~\eqref{eq:dist_sum_gen}, we get
	%
	%
	%	\begin{subequations}
	\begin{align}
	f_Z(z) = \frac{1}{2}N\beta^{-N}(2x+2\beta-z)^{N-1},
	\end{align}
	for $2x+\beta<z<2(x+\beta)$ and
	\begin{align}
	f_Z(z) = -\frac{1}{2}N\beta^{-N}\frac{(z-2x)^{N}}{2x -z},
	\end{align}
	for $2x<z\leq2x+\beta$ and zero otherwise.
	%	\end{subequations}
	%
	%	\begin{equation}
	%	f_Z(z) =
	%	\begin{cases}
	%	\frac{1}{2}N^{-N}(2x+2a_u-z)^{N-1}& \parbox[t]{3cm}{\raggedright $2x+a_u<z<2(x+a_u)$}\\
	%	\frac{1}{2}Na_u^{-N}(z-2x)^{N-1} & \parbox[t]{4cm}{\raggedright $2x<z\leq2x+a_u$}\\
	%	0 & \parbox[t]{4cm}{\raggedright Otherwise.}\\
	%	\end{cases}
	%	\end{equation}
	%
	%
	It can be shown that
	\begin{align}
	\E(f_Z(z)) = 2x + \beta.
	\end{align}
\end{subequations}
Hence, noting that $\beta$ is known, the function $g(T_1(y_{1:N}),T_2(y_{1:N}))$ that gives an unbiased estimator should be of the form of
\begin{subequations}
	\begin{align}
	\hat{x}_{\mathrm{MVU}}^{\mathcal{U}}(y_{1:N},\beta) &= g(T_1(y_{1:N}),T_2(y_{1:N}))\nonumber\\
	&= \frac{1}{2}(y_{(1)}+ y_{(N)}) - \frac{\beta}{2}.
	\end{align}
	The MSE of the MVU estimator is given by
	\begin{align}
	\MSE\left(\hat{x}_{\mathrm{MVU}}^{\mathcal{U}}(y_{1:N},\beta) \right) = \frac{\beta^2}{2N(N+3)+4}.\label{eq:mse_mvu_unif_known}
	\end{align}
\end{subequations}
Comparing to~\eqref{eq:mse_blue_uniform}, %\[\MSE\left(\hat{x}_{\mathrm{BLUE}}^{\mathcal{U}}(\bm{y},\beta) \right) \propto N\cdot\MSE\left(\hat{x}_{\mathrm{MVU}}^{\mathcal{U}}(\bm{y},\beta) \right)\] and 
the order statistics based MVU estimator outperforms the BLUE one order of magnitude. 

\subsubsection{Unknown hyper parameter $\beta$}\label{subsec:unknown_hyper_parameter_uniform}
In this case, the MVU estimators for the parameter vector $\Psi=[x,\beta]^\top$ can be derived from sufficient statistics~\eqref{eq:uniform_ss},
\begin{align}
\hat{\Psi}=g(T(y_{1:N})),\quad \mathrm{s.t.}\quad \E\left(g\left(T(y_{1:N})\right)\right) = \Psi.
\end{align}
In this case, we have
\begin{align}
\E(T(y_{1:N})) = \begin{bmatrix}
x+\frac{\beta}{N+1}\\\\x+\frac{N\beta}{N+1}
\end{bmatrix}
\label{eq:uniform_unknown_E_gen}
\end{align}
To find the transformation that makes~\eqref{eq:uniform_unknown_E_gen} unbiased, we define
\begin{subequations}
	\begin{align}
	g(T(y_{1:N}))=\begin{bmatrix}\frac{1}{N-1}\left(NT_1(y_{1:N})-T_2(y_{1:N})\right)\\\\	\frac{N+1}{N-1}\left(T_2(y_{1:N})-T_1(y_{1:N})\right)\end{bmatrix}
	\end{align}
	that gives
	\begin{align}
	\E\left(g(T(y_{1:N}))\right) = \begin{bmatrix}x\\\beta\end{bmatrix}.
	\end{align}
\end{subequations}
Finally, the MVU estimator of $x$ when the hyper parameter $\beta$ is unknown is given by
\begin{subequations}
	\begin{align}
	\hat{x}_{\mathrm{MVU}}^{\mathcal{U}}(y_{1:N}) = \frac{N}{N-1}y_{(1)} - \frac{1}{N-1} y_{(N)}.
	\end{align}
	and its MSE is
	\begin{align}
	\MSE\left(\hat{x}_{\mathrm{MVU}}^{\mathcal{U}}(y_{1:N})\right) = \frac{N\beta^2}{(N+2)(N^2-1)}.
	\end{align}
\end{subequations}
This is naturally slightly larger than~\eqref{eq:mse_mvu_unif_known} for finite $N$.
%%%%%%%%%%%%%%%%%%%%%%%%%%%%%%%%%%%%
%---------------------------SECTION----------------------------
%%%%%%%%%%%%%%%%%%%%%%%%%%%%%%%%%%%%
\section{Distributions in the exponential family} \label{sec:distributioons_in_the_exponential_family}
% Beta and Gamma distributions, in their general form, do not give a closed-form solution for the bias and variance. However, there are other continuous non-negative distributions for which the analytical solutions exists. In this section we investigate such distributions and show that the marginal distribution of the minimum order statistic has the same "form" of distribution as the noise.
% Multiple distributions with positive support have been tested.

The exponential family of probability distributions, in their most general form, is defined by
\begin{align}
f(y;\theta) = h(y)g(\theta)\exp\left\{A(\theta)\cdot T(y)\right\},
\end{align}
where $\theta$ is the parameter of the distribution, and $h(y)$, $g(\theta)$, $A(\theta)$, and $T(y)$ are all known functions. In this section, we only consider some example distributions of this family and show that the minimum order statistic estimator gets the same form of distribution as the noise distribution but with modified parameters. For the selected distributions, if possible, MVU estimators for both cases of known and unknown hyperparameter  are derived. Otherwise, unbiased estimators with less variance than BLUE are proposed.
%Note that, in all distributions of the exponential family, the support of $f(y;\theta)$ does not depend on $\theta$. This excludes Pareto distribution, that will be discussed in the next section, from the exponential family. 
 %<<<<<<<<<<<<<<<<<<<<<<<<<<<<<<<<<<<<<<<<<<<<<<<<
%------------------------SUBSECTION--------------------------
%<<<<<<<<<<<<<<<<<<<<<<<<<<<<<<<<<<<<<<<<<<<<<<<<  
\subsection{Exponential distribution} \label{subsec:exponential_distribution}
Exponential distributions are members of the gamma family with shape parameter 1; strongly skewed with no left sided tail ($y_k\in[x,\infty]$). Let $\beta>0$ denote the scale parameter, the PDF of an exponential distribution is then given by
\begin{subequations}\label{eq:exponential}
	\begin{align}
	f^{\mathrm{Exp}}(y_k;x,\beta)=\left\{\begin{matrix}
	\frac{1}{\beta}\exp(-\frac{(y_k-x)}{\beta})&y_k\geq x, \\ 
	0&y_k< x. 
	\end{matrix}\right.
	\label{eq:exponential_pdf}
	\end{align}
		and the CDF, for $y\geq x$, is given by
		\begin{align}
		F^{\mathrm{Exp}}(y_k;x,\beta)=1-\exp(-\frac{(y_k-x)}{\beta}).
		\label{eq:exponential_cdf}
		\end{align}
\end{subequations}

For the BLUE estimator~\eqref{eq:mean_estimator}, from the properties of exponential distribution, we have
\begin{align}
\hat{x}_{\mathrm{BLUE}}^{\mathrm{Exp}}(y_{1:N},\beta) &= \frac{1}{N}\sum_{k=1}^{N}y_k - \beta,
&\quad 
\MSE(\hat{x}_{\mathrm{BLUE}}^{\mathrm{Exp}}) &= \frac{\beta^2}{N}.
\end{align}

Substituting~\eqref{eq:exponential} into~\eqref{eq:density_order}, the marginal density of the $k$:th order statistic is given by
\begin{align}
f^{\mathrm{Exp}}_{(k,N)}(y;x,\beta) = \frac{N}{\beta}\begin{pmatrix}N-1\\k-1\end{pmatrix}\left(1-\exp(-\frac{(y-x)}{\beta})\right)^{k-1}\exp\left(-\frac{(N-k+1)(y-x)}{\beta}\right).
\label{eq:exponential_order}
\end{align}

The first order statistic density is then given by letting $k=1$ in~\eqref{eq:exponential_order} that results in another exponential distribution,
\begin{align}
f^{\mathrm{Exp}}_{(1,N)}(y;x,\beta) = f^{\mathrm{Exp}}(y;x,\bar{\beta}),
\end{align}
where $\bar{\beta}=\frac{\beta}{N}$. Hence, the MSE of the minimum order statistics estimator is given by
\begin{align}
\MSE\left(\hat{x}_{\mathrm{min}}^{\mathrm{Exp}}(y_{1:N})\right) = \frac{2\beta^2}{N^2}.
\end{align}
In order to find the MVU estimator, we re-write the PDF as
\begin{align}
f(y_{1:N};x,\beta) &= \frac{1}{\beta^N}\exp\left[-\frac{1}{\beta}\sum_{k=1}^{N}y_k\right]\exp\left[-\frac{N}{\beta}x\right]\times\sigma(y_{(1)} - x).
\label{eq:exponential_pdf_2}
\end{align}
\subsubsection{Known hyper parameter $\beta$}
In case of the known hyper parameter $\beta$, the Neyman-Fisher factorization of PDF~\eqref{eq:exponential_pdf_2} gives
\begin{subequations}
	\begin{align}
	T(y_{1:N}) &= y_{(1)}\\
	h(y_{1:N}) &=   \frac{1}{\beta^N}\exp\left[-\frac{1}{\beta}\sum_{k=1}^{N}y_k\right].
	\label{eq:exponential_known_ss}
	\end{align}
\end{subequations}
The MVU estimator can then be obtained from a transformation of the minimum order statistic that makes it an unbiased estimator. Finally, in case of exponential noise with known hyper parameter of the distribution, the MVU estimator and its MSE are given by
\begin{subequations}
	\begin{align}
	\hat{x}_{MVU}^{\mathrm{Exp}}(y_{1:N},\beta) &= y_{(1)} - \frac{\beta}{N}
	\\
	\MSE\left(\hat{x}_{MVU}^{\mathrm{Exp}}(y_{1:N},\beta)\right) &= \frac{\beta^2}{N^2}.
	\end{align}
\end{subequations}
\subsubsection{Unknown hyper parameter $\beta$}
If the hyper parameter $\beta$ is unknown, the factorization gives
\begin{subequations}
	\begin{align}
	T(y_{1:N}) = \begin{bmatrix}
	y_{(1)}\\ \sum_{k=1}^{N} y_k
	\end{bmatrix} = \begin{bmatrix}
	T_1(y_{1:N}) \\ T_2(y_{1:N})
	\end{bmatrix}.
	\label{eq:uniform_unknown_ss}
	\end{align}
	Noting that sum of exponential random variables results in a Gamma distribution, we have $T_2(y_{1:N})\sim\Gamma(N,\beta)$. Hence,
	\begin{align}
	\E(T(y_{1:N})) = \begin{bmatrix}
	x+\frac{\beta}{N}\\\\N(x+\beta)
	\end{bmatrix}.
	\label{eq:exponential_unknown_E_gen}
	\end{align}
\end{subequations}

Following the same line of reasoning as in Section~\ref{subsec:unknown_hyper_parameter_uniform}, the unbiased estimator is given by the transformation
\begin{subequations}
	\begin{align}
	g(T(y_{1:N}))=\begin{bmatrix}\frac{1}{N-1}\left(NT_1(y_{1:N})-\frac{1}{N}T_2(y_{1:N})\right)\\\\	\frac{1}{N-1}\left(T_2(y_{1:N})-NT_1(y_{1:N})\right)\end{bmatrix},
	\end{align}
	that gives
	\begin{align}
	\E\left(g(T(y_{1:N}))\right) = \begin{bmatrix}x\\\beta	\end{bmatrix}.
	\end{align}
\end{subequations}
Finally, the MVU estimator when the hyper parameter $\beta$ is unknown, is given by
\begin{subequations}
	\begin{align}
	\hat{x}_{\mathrm{MVU}}^{\mathrm{Exp}}(y_{1:N}) &= \frac{N}{N-1}y_{(1)} - \frac{1}{N(N-1)} \sum_{k=1}^{N} y_k \nonumber\\&= \frac{N}{N-1}y_{(1)}- \frac{1}{N-1} \bar{y},
	\end{align}
	where $\bar{y}$ is the sample mean. Assuming that $N$ is large $\min_k y_k$ and $\bar{y}$ are independent and the MSE of the estimator, asymptotically, is given by
	\begin{align}
	\MSE\left(\hat{x}_{\mathrm{MVU}}^{\mathrm{Exp}}(y_{1:N})\right) = \frac{\beta^2(N+1)}{N(N-1)^2}.
	\end{align}
\end{subequations}
%
%
%<<<<<<<<<<<<<<<<<<<<<<<<<<<<<<<<<<<<<<<<<<<<<<<<
%------------------------SUBSECTION--------------------------
%<<<<<<<<<<<<<<<<<<<<<<<<<<<<<<<<<<<<<<<<<<<<<<<<
\subsection{Rayleigh distribution} \label{subsec:rayleigh_distribution}
One generalization of the exponential distribution is obtained by parameterizing in terms of both a scale parameter $\beta$ and a shape parameter $\alpha$. Rayleigh distribution is a special case obtained by setting $\alpha=2$
\begin{subequations}\label{eq:rayleigh}
	\begin{align}
	f^{\mathrm{Rayleigh}}(y_k;x,\beta)=\left\{\begin{matrix}
	\frac{y_k-x}{\beta^2}\exp(-\frac{(y_k-x)^2}{2\beta^2})&y_k> x, \\ 
	0&y_k\leq x. 
	\end{matrix}\right.
	\label{eq:rayleigh_pdf}
	\end{align}
	and the CDF, for $y_k>x$ is given by
	\begin{align}
	F^{\mathrm{Rayleigh}}(y_k;x,\beta)=1-\exp(-\frac{(y_k-x)^2}{2\beta^2}).
	\label{eq:rayleigh_cdf}
	\end{align}
\end{subequations}
Hence, the BLUE estimator~\eqref{eq:mean_estimator}, becomes
\begin{subequations}
	\begin{align}
	\hat{x}_{\mathrm{BLUE}}^{\mathrm{Rayleigh}}(y_{1:N},\beta) &= \frac{1}{N}\sum_{k=1}^{N}y_k - \sqrt{\frac{\pi}{2}}\beta,
	\\
	\MSE\left(\hat{x}^{\mathrm{Rayleigh}}_{\mathrm{BLUE}}(y_{1:N},\beta)\right) &= \frac{(4-\pi)\beta^2}{2N}.
	\end{align}
\end{subequations}
The marginal density of the $k$:th order statistic is given by
\begin{align}
f^{\mathrm{Rayleigh}}_{(k,N)}(y;x,\beta) =\left\{\begin{matrix}
\frac{Ny}{\beta^2}\begin{pmatrix}N-1\\k-1\end{pmatrix}\left(1-\exp(-\frac{(y-x)^2}{2\beta^2})\right)^{k-1}\exp\left(-\frac{(N-k+1)(y-x)^2}{2\beta^2}\right)&y> x, \\ 
0&y\leq x. 
\end{matrix}\right. 
\label{eq:rayleigh_order}
\end{align}
Hence, the minimum order statistics density also is Rayleigh distributed 
\begin{align}
f^{\mathrm{Rayleigh}}_{(1,N)}(y;x,\beta) = f^{\mathrm{Rayleigh}}(y;x,\bar{\beta}),
\end{align}
where $\bar{\beta}=\frac{\beta}{\sqrt{N}}$.  The MSE of the minimum order statistics is given by
\begin{align}
\MSE\left(\hat{x}^{\mathrm{Rayleigh}}_{\mathrm{min}}(y_{1:N})\right) = \frac{2\beta^2}{N}.
\end{align}
The joint PDF of $N$ independent observations $y_{1:N}$ is given by
\begin{subequations}\label{eq:rayleigh_pdf_2}
	\begin{align}
	f(y_{1:N};x,\beta) &= \frac{\prod_{k=1}^{N}(y_k-x)}{\beta^{2N}}\exp\left[\sum_{k=1}^{N}-\frac{(y_k-x)^2}{2\beta^2}\right]\sigma(y_{(1)}- x).
	\label{eq:rayleigh_pdf_2_1}
	\end{align}
	Noting that
	\begin{align}
	\sum_{k=1}^{N}(y_k-x)^2 = \sum_{k+1}^{N}(y_k)^2 - 2x\sum_{k=1}^{N}y_k+Nx^2,
	\label{eq:rayleigh_pdf_2_2}
	\end{align}
	the PDF becomes
	\begin{align}
	f(y_{1:N};x,\beta) &= \beta^{-2N}\prod_{k=1}^{N}(y_k-x)\exp\left[\frac{-1}{2\beta^2}\sum_{k=1}^{N}y_k^2\right]\nonumber\\&\times\exp\left[-\frac{Nx^2}{2\beta^2}\right]\exp\left[\frac{x}{\beta^2}\sum_{k=1}^{N}y_k\right]\sigma(y_{(1)} - x).
	\label{eq:rayleigh_pdf_2_3}
	\end{align}
\end{subequations}
\subsubsection{Known hyper parameter $\beta$}
Since~\eqref{eq:rayleigh_pdf_2_3} cannot be factorized in the form of $f(y_{1:N};x,\beta) = g(T(y_{1:N}),x)h(y_{1:N})$, the RBLS theorem cannot be used. Hence, even if an MVU estimator exists for this problem, we may not be able to find it. Thus, in case of Rayleigh-distributed measurement noise, we propose unbiased estimators based on order statistics. 

If the hyper parameter of the distribution is known, the unbiased order statistic based estimator $\hat{x}^{\mathrm{Rayleigh}}(y_{1:N},\beta)$ is then given by,
\begin{subequations}
	\begin{align}
	\hat{x}^{\mathrm{Rayleigh}}(y_{1:N},\beta) &= y_{(1)} - \frac{\sqrt{\pi}\beta}{\sqrt{2N}},
	\\ 
	\MSE\left(\hat{x}^{\mathrm{Rayleigh}}(y_{1:N},\beta)\right) &=\frac{(4-\pi)\beta^2}{2N}.
	\end{align}
\end{subequations}
which has the same variance as the BLUE estimator.
\subsubsection{Unknown hyper parameter $\beta$}
 In case of unknown hyper parameters, as for the known case, no factorization that enables us to use the RBLS theorem can be found. In this case, we propose the following unbiased estimator
\begin{align}
\hat{x}^{\mathrm{Rayleigh}}(y_{1:N}) &= \frac{\sqrt{N}}{\sqrt{N}-1}y_{(1)} - \frac{1}{N(\sqrt{N}-1)}\sum_{k=1}^{N}y_k\nonumber\\&= \frac{1}{\sqrt N-1}(\sqrt{N}y_{(1)} - \bar{y}).
\end{align}
Asymptotically, for large $N$, the sample mean and minimum order statistic are independent and the estimator MSE is given by
\begin{align}
\MSE\left(\hat{x}^{\mathrm{Rayleigh}}(y_{1:N})\right)=\frac{(1+N)(4-\pi)\beta^2}{2N(\sqrt{N}-1)^2}.
\label{eq:minimum_orde_unknown_rayleigh}
\end{align}
%
%
%<<<<<<<<<<<<<<<<<<<<<<<<<<<<<<<<<<<<<<<<<<<<<<<<
%------------------------SUBSECTION--------------------------
%<<<<<<<<<<<<<<<<<<<<<<<<<<<<<<<<<<<<<<<<<<<<<<<<
\subsection{Weibull distribution} \label{subsec:weibull_distribution}
Weibull distribution is a generalization of the Rayleigh, distribution that is parameterized by two parameters--scale parameter $\beta$ and shape parameter $\alpha>0$. In fact Weibull distribution is obtained by relaxing the assumption $\alpha=2$ in the Rayleigh distribution and its density function is given by
\begin{subequations}\label{eq:weibull}
	\begin{align}
	f^{\mathrm{Weibull}}(y_k;x,\beta,\alpha)=\left\{\begin{matrix}
	\frac{\alpha}{\beta}\left(\frac{y_k-x}{\beta}\right)^{\alpha-1}\exp(-(\frac{y_k-x}{\beta})^\alpha)&y_k> x, \\ 
	0&y_k\leq x. 
	\end{matrix}\right.
	\label{eq:weibull_pdf}
	\end{align}
	and the CDF, for $y_k\geq x$ is given by
	\begin{align}
	F^{\mathrm{Weibull}}(y_k;x,\beta,\alpha)=1-\exp(-(\frac{y_k-x}{\beta})^\alpha).
	\label{eq:weibull_cdf}
	\end{align}
	The BLUE estimator, in case of Weibull-distributed measurement noises is given by
\begin{align}
\hat{x}_{\mathrm{BLUE}}^{\mathrm{Weibull}}(y_{1:N},\beta,\alpha) &= \frac{1}{N}\sum_{k=1}^{N}y_k - \beta\Gamma(1+\frac{1}{\alpha}),
&\quad 
\MSE(\hat{x}_{\mathrm{BLUE}}^{\mathrm{Weibull}}(y_{1:N},\beta,\alpha)) &= \frac{\beta^2}{N}\left[\Gamma(\frac{\alpha+2}{\alpha})-\left(\Gamma(\frac{\alpha+1}{\alpha})\right)^2\right].
\end{align}
\end{subequations}
The marginal density of the $k$:th order statistic is given by
\begin{align}
f^{\mathrm{Weibull}}_{(k,N)}(y;x,\beta,\alpha) = \frac{N\alpha}{\beta}\begin{pmatrix}N-1\\k-1\end{pmatrix}(\frac{y-x}{\beta})^{\alpha-1}\left(1-\exp^{-(\frac{y-x}{\beta})^\alpha}\right)^{k-1}\exp\left(-(N-k+1)(\frac{y-x}{\beta})^\alpha\right).
\label{eq:weibull_order}
\end{align}
Hence, the first order statistic density in case of $e_k\sim\mathrm{Weibull}(\beta,\alpha)$, is another Weibull distribution,
\begin{align}
f^{\mathrm{Weibull}}_{(1,N)}(y;x,\beta,\alpha) = f^{\mathrm{Weibull}}(y;x,\bar{\beta},\alpha),
\end{align}
where $\bar{\beta}=\sqrt[-\alpha]{N}\beta$. This gives the MSE of the minimum order statistic estimator as
\begin{align}
\MSE(\hat{x}_{\mathrm{min}}^{\mathrm{Weibull}}(y_{1:N})) = \beta^2N^{\frac{-2}{\alpha}}\Gamma(\frac{\alpha+2}{\alpha})
\end{align}
Given $N$ independent observations, the joint density is given by
	\begin{align}
f^{\mathrm{Weibull}}(y_{1:N};x,\beta,\alpha)=(\frac{\alpha}{\beta})^N\prod_{k=1}^{N}\left(\frac{y_k-x}{\beta}\right)^{\alpha-1}\exp(-\sum_{k=1}^{N}(\frac{y_k-x}{\beta})^\alpha)\sigma(y_{(1)} - x)
\label{eq:weibull_pdf_joint}
\end{align}
Since~\eqref{eq:weibull_pdf_joint} cannot be factorized using Neyman-Fisher factorization, RBLS is not applicable. Additionally, in this case, it is not possible to find an unbiased estimator when the hyper parameters $\alpha$ and $\beta$ are unknown. In case of known hyper parameters, the unbiased minimum order statistic estimator, however, can be computed.  The unbiased estimator based on minimum order statistic is given by,
	\begin{align}
	\hat{x}^{\mathrm{Weibull}}(y_{1:N},\beta,\alpha) &= y_{(1)}- \beta N^{-\frac{1}{\alpha}}\Gamma(1+\frac{1}{\alpha}),
	&\nonumber\\ 
	\MSE(\hat{x}^{\mathrm{Weibull}}(y_{1:N},\beta,\alpha)) &=\beta^2N^{\frac{-2}{\alpha}}\left[\Gamma(\frac{\alpha+2}{\alpha})-\left(\Gamma(\frac{\alpha+1}{\alpha})\right)^2\right].
	\end{align}
An order-statistics-based unbiased estimator with unknown hyper parameters of the distribution could not be obtained.
\section{Other Distributions}\label{sec:other_distributions}
In this section, we further study the location estimation problem for two other noise distributions. In the rest, the Pareto distribution with positive support is first studied followed by the mixture of uniform and normal distribution.
\subsection{Pareto distribution} 
Let the scale parameter $\beta$ (necessarily positive) denote the minimum possible value of $y_k$, and $\alpha>0$ denote the shape parameter. The Pareto Type I distribution is characterized by $\beta$ and $\alpha$
\begin{subequations}\label{eq:pareto}
	\begin{align}
	f^{\mathrm{Pareto}}(y_k;x,\beta,\alpha)=\left\{\begin{matrix}
	\alpha\beta^\alpha (y_k-x)^{-(\alpha+1)}&y_k\geq x+\beta, \\ 
	0&y_k< x+\beta. 
	\end{matrix}\right.
	\label{eq:pareto_pdf}
	\end{align}
	and the CDF is given by
	\begin{align}
	F^{\mathrm{Pareto}}(y_k;x,\beta,\alpha)=1-\left(\frac{\beta}{y-x}\right)^\alpha.
	\label{eq:pareto_cdf}
	\end{align}
\end{subequations}
For the BLUE we get,
\begin{subequations}
	\begin{align}
	\hat{x}_{\mathrm{BLUE}}^{\mathrm{Pareto}}(y_{1:N},\beta,\alpha) &= \frac{1}{N}\sum_{k=1}^{N}y_k - \frac{\alpha\beta}{\alpha-1},&\quad \alpha>1
	\\
	\MSE(\hat{x}_{\mathrm{BLUE}}^{\mathrm{Pareto}}(y_{1:N},\beta,\alpha)) &=\frac{\alpha\beta^2}{N(\alpha-1)^2(\alpha-2)},&\quad \alpha>2
	\end{align}
\end{subequations}
The RBLS theorem cannot be used in case Pareto-distributed noises. We provide an unbiased estimator using minimum order statistics and its variance. The marginal density of the $k$:th order statistic, for $y\geq x+\beta$ is given by
\begin{align}
f^{\mathrm{Pareto}}_{(k,N)}(y;x,\beta,\alpha) = N\alpha\beta^\alpha (y-x)^{-(\alpha+1)}\begin{pmatrix}N-1\\k-1\end{pmatrix}\left(1-(\frac{\beta}{y-x})^\alpha\right)^{k-1}\left(\frac{\beta}{y-x}\right){\alpha(N-k)}.
\label{eq:pareto_order}
\end{align}
The minimum order statistic has the same form of distribution
\begin{align}
f^{\mathrm{Pareto}}_{(1,N)}(y;\beta,\alpha) = f^{\mathrm{Pareto}}(y;\beta,\bar{\alpha}),
\end{align}
where $\bar{\alpha}=N\alpha$. The MSE of the minimum order statistic estimator is 
\begin{align}
\MSE(\hat{x}_{\mathrm{min}}^{\mathrm{Pareto}}(y_{1:N})) =  \frac{N\alpha\beta^2}{N\alpha-2}
\end{align}
The unbiased estimator is thus given by,
\begin{subequations}
	\begin{align}
	\hat{x}^{\mathrm{Pareto}}(y_{1:N},\beta,\alpha) &= y_{(1)} - \frac{N\alpha\beta}{N\alpha-1},&\quad N\alpha>1
	\\
	\MSE(\hat{x}^{\mathrm{Pareto}}(y_{1:N},\beta,\alpha)) &=\frac{N\alpha\beta^2}{(N\alpha-1)^2(N\alpha-2)},&\quad N\alpha>2
	\end{align}
\end{subequations}

No unbiased estimator for unknown hyper parameter case could be found for Pareto distribution. 
%%%%%%%%%%%%%%%%%%%%%%%%%%%%%%%%%%%%
%---------------------------SECTION----------------------------
%%%%%%%%%%%%%%%%%%%%%%%%%%%%%%%%%%%%
\section{Mixture of Normal and Uniform Noise Distribution} 
Suppose the error is distributed as \[ e_k\sim \alpha\mathcal{N}(0,\sigma^2) + (1-\alpha) \mathcal{U}[0,\beta],\] where $\alpha$ is the mixing probability of the mixture distribution. Define $f^{\mathrm{\mathcal{U},\mathcal{N}}}(y_k)$ as the probability density function of the considered mixture distribution given by
\begin{align}\label{eq:mixture_pdf}
f&^{\mathrm{\mathcal{U},\mathcal{N}}}(y_k;x,\alpha,\sigma^2,\beta) =\nonumber\\
&\begin{cases}
\begin{aligned}
& \frac{a}{\sqrt{2\pi\sigma^2}}\exp\left[-\frac{(y_k-x)^2}{2\sigma^2}\right] +\frac{1-a}{\beta}		\end{aligned} & \parbox[t]{4cm}{\raggedright $0\leq y_k-x\leq \beta$}\\
\frac{a}{\sqrt{2\pi\sigma^2}}\exp\left[-\frac{(y_k-x)^2}{2\sigma^2}\right] & \parbox[t]{4cm}{\raggedright Otherwise.}\\
\end{cases} 
\end{align}
The BLUE, in case of the mixture of normal and uniform measurement noises is given by
\begin{subequations}
	\begin{align}
	\hat{x}_{\mathrm{BLUE}}^{\mathrm{\mathrm{\mathcal{U},\mathcal{N}}}}(y_{1:N},\alpha,\beta,\sigma^2) &= \frac{1}{N}\sum_{k=1}^{N}y_k - \frac{\beta(1-\alpha)}{2},
	\\
	\MSE\left(\hat{x}_{\mathrm{BLUE}}^{\mathrm{\mathrm{\mathcal{U},\mathcal{N}}}}(y_{1:N},\alpha,\beta,\sigma^2)\right) &=\nonumber\\ &\frac{\beta^2  \left(1+(2-3\alpha)\alpha\right)+12\alpha\sigma^2}{12N}.
	\end{align}
\end{subequations}

Noting that at $y_k-x=0$ contributions of the uniform distribution and the mean (mode) of the normal distribution are added together,~\eqref{eq:mixture_pdf}  is maximized at this point. The order statistics PDF for $0\leq y-x\leq \beta$ is given by
\begin{align}
f_{(k,N)}^{\mathrm{\mathrm{\mathcal{U},\mathcal{N}}}}&(y;\alpha,\beta,\sigma^2,x,k) =\nonumber\\ &N\begin{pmatrix}N-1\\k-1\end{pmatrix}\left(\frac{\alpha\exp(-\frac{(y-x)^2}{2\sigma^2})}{\sqrt{2\pi\sigma^2}}+\frac{1-\alpha}{\beta}\right)\nonumber\\
&\times\left(\frac{(1-\alpha)(y-x)}{\beta}+\frac{\alpha}{2}(1+\mathrm{Erf}\left[\frac{y-x}{\sqrt{2\sigma^2}}\right])\right)^{k-1}\nonumber\\
&\times\left(1+\frac{(\alpha-1)(y-x)}{\beta}-\frac{\alpha}{2}(1+\mathrm{Erf}\left[\frac{y-x}{\sqrt{2}\sigma}\right])\right)^{N-k},
\end{align}
where $\mathrm{Erf}(\cdot) = \frac{2}{\sqrt{\pi}}\int_{0}^{\cdot}e^{-t^2}dt$ is the error function. In order to find the best order statistic estimator, we maximize the likelihood function $\ell(k\mid y=x,a,\beta,\sigma^2)$
\begin{subequations}
	\begin{align}
	\ell(k\mid y=x,\alpha,\beta,\sigma^2) &= N\begin{pmatrix}N-1\\k-1\end{pmatrix}2(2-\alpha)^{-k}(1-\frac{\alpha}{2})^N\alpha^{k-1}\nonumber\\&\times\left(\frac{1-\alpha}{\beta}+\frac{\alpha}{\sqrt{2\pi}\sigma}\right).
	\label{eq:best_order_likelihood_general}
	\end{align}
	Noting that $\left(\frac{1-\alpha}{\beta}+\frac{\alpha}{\sqrt{2\pi}\sigma}\right)$ is always positive and independent of $k$, we extract it from the likelihood function. Simplifying~\eqref{eq:best_order_likelihood_general} by means of manipulating the terms, we get

		\begin{align}
		2(2-\alpha)^{-k}&=2^{1-k}(1-\frac{\alpha}{2})^{-k},\\
		\alpha^{k-1}&=(2\frac{\alpha}{2})^{k-1}=2^{k-1}(\frac{\alpha}{2})^{k-1}.
		\end{align}

	the likelihood function to be maximized can be re-written as
	\begin{align}
	\ell(k\mid y=x,\alpha,,\sigma^2) \propto \begin{pmatrix}N-1\\k-1\end{pmatrix}(\frac{\alpha}{2})^{k-1}(1-\frac{\alpha}{2})^{N-k}.
	\label{eq:best_order_likelihood_modified}
	\end{align}
\end{subequations}
In order to find the maximum likelihood estimate $\hat{k} = \argmax_k \ell(k\mid y-x=0)$, we note that~\eqref{eq:best_order_likelihood_modified} is a binomial distribution with probability of success $\frac{\alpha}{2}$. Hence, the maximum of the function is given at the mode of the distribution, 
\begin{align}
\hat{k}=\lfloor \frac{N\alpha}{2}\rfloor+1\quad \mathrm{or}\quad \lceil\frac{N\alpha}{2}\rceil.
\end{align}
This gives the best order statistic estimator for the case when noise is a mixture of normal and uniform distribution as
\begin{align}
\hat{x}^{\mathrm{\mathrm{\mathcal{U},\mathcal{N}}}}(y_{1:N},\alpha)= y_{(\hat{k})}.
\end{align}
\begin{table}[ht]
	\centering
	\caption{Bias and MSE of minimum order statistics estimators $\hat{x}_{\mathrm{min}}^{p_e}$.}
	\begin{tabular}{c|c|c}
		\hline
		distribution                     & bias                                                   & MSE                                                      \\ \hline
		$\mathcal{U}[0,\beta]$               & $\frac{\beta}{N+1}$                                        & $\frac{2\beta^2}{(N+1)(N+2)}$                            \\[2mm] \hline
		$\mathrm{Exp}(\beta)$            & $\frac{\beta}{N}$                                      & $\frac{2\beta^2}{N^2}$                                   \\[2mm] \hline
		$\mathrm{Rayleigh}(\beta)$       & $\frac{\sqrt{\pi}\beta}{\sqrt{2N}}$                    & $\frac{2\beta^2}{N}$                                     \\[2mm] \hline
		$\mathrm{Weibull}(\beta,\alpha)$ & $\beta N^{-\frac{1}{\alpha}}\Gamma(1+\frac{1}{\alpha})$ & $\beta^2N^{-\frac{2}{\alpha}}\Gamma(1+\frac{2}{\alpha})$ \\[2mm] \hline
		$\mathrm{Pareto}(\beta,\alpha)$  & $\frac{N\alpha\beta}{N\alpha -1}$                      & $\frac{N\alpha\beta^2}{N\alpha -2}$                      \\ \hline
	\end{tabular}
	\label{tbl:estimators_biased}
\end{table}
\begin{table}[t]
	\centering
	\caption{Estimators and their MSEs derived for multiple noise distributions.}
	\resizebox{\textwidth}{!}{
		\begin{tabular}{c|l|l|cl}
			\hline
			\begin{tabular}[c]{@{}c@{}}noise\\ distribution\end{tabular} &
			 \multicolumn{2}{c|}{estimator}                                                                                            & \multicolumn{2}{c}{MSE}                                                                                                         \\  \hline
		\multirow{3}{*}{\vspace{-6mm}$\mathcal{U}[0,\beta]$}                          & \multicolumn{2}{l|}{$\hat{x}_\mathrm{BLUE}^{\mathcal{U}}(y_{1:N},a_n) = \frac{1}{N}\sum_{k=1}^{N} y_k - \frac{\beta}{2}$}                                          &   \multicolumn{2}{c}{$\frac{\beta^2}{12N}$}                                                                                                 \\[2mm] 
		& \multicolumn{2}{l|}{$\hat{x}_\mathrm{MVU}^{\mathcal{U}}(y_{1:N},a_n)  = \frac{1}{2}(y_{(1)} + y_{(N)}) - \frac{\beta}{2}$}                     & \multicolumn{2}{c}{$\frac{\beta^2}{2N(N+3)+4}$}                                                                                           \\[2mm] 
		& \multicolumn{2}{l|}{$\hat{x}_\mathrm{MVU}^{\mathcal{U}}(y_{1:N})  = \frac{N}{N-1}y_{(1)} - \frac{1}{N-1} y_{(N)}$}                   & \multicolumn{2}{c}{$\frac{N\beta^2}{(N+2)(N^2-1)}$}                                                                                        \\[2mm]   \hline
					\multirow{3}{*}{\vspace{-6mm}$\mathrm{Exp}(\beta)$}                       & \multicolumn{2}{l|}{$\hat{x}_\mathrm{BLUE}^{\mathrm{Exp}}(y_{1:N},\beta)  = \frac{1}{N}\sum_{k=1}^{N} y_k -\beta$}                                                 & \multicolumn{2}{c}{$\frac{\beta^2}{N}$}                                                                                             \\[2mm]
		& \multicolumn{2}{l|}{$\hat{x}_\mathrm{MVU}^{\mathrm{Exp}}(y_{1:N},\beta) = y_{(1)} - \frac{\beta}{N}$}                                   & \multicolumn{2}{c}{$\frac{\beta^2}{N^2}$}                                                                                           \\[2mm]
		& \multicolumn{2}{l|}{$\hat{x}_\mathrm{MVU}^{\mathrm{Exp}}(y_{1:N}) = \frac{N}{N-1}y_{(1)}- \frac{\sum_{k=1}^{N} y_k}{N(N-1)}$}           & \multicolumn{2}{c}{$\frac{(N+1)\beta^2}{N(N-1)^2}$}                                                                                 \\[2mm] \hline
				\multirow{3}{*}{\vspace{-6mm}$\mathrm{Rayleigh}(\beta)$}                  & \multicolumn{2}{l|}{$\hat{x}_\mathrm{BLUE}^{\mathrm{Rayleigh}}(y_{1:N},\beta) = \frac{1}{N}\sum_{k=1}^{N} y_k -\sqrt{\frac{\pi}{2}}\beta$}                             & \multicolumn{2}{c}{$\frac{(4-\pi)\beta^2}{2N}$}                                                                                     \\[2mm]
			& \multicolumn{2}{l|}{$\hat{x}^{\mathrm{Rayleigh}}(y_{1:N},\beta) = y_{(1)} - \frac{\sqrt{\pi}\beta}{\sqrt{2N}}$}                              & \multicolumn{2}{c}{$\frac{(4-\pi)\beta^2}{2N}$}                                                                                      \\[2mm]
			& \multicolumn{2}{l|}{$\hat{x}^{\mathrm{Rayleigh}}(y_{1:N}) = \frac{\sqrt{N}}{\sqrt{N}-1}y_{(1)} - \frac{\sum_{k=1}^{N} y_k}{N(\sqrt{N}-1)}$}                                  & \multicolumn{2}{c}{$\frac{(1+N)(4-\pi)\beta^2}{2N(\sqrt{N}-1)^2}$}                       
			\\[2mm] \hline
			\multirow{2}{*}{\vspace{-4mm}$\mathrm{Weibull}(\beta,\alpha)$}            & \multicolumn{2}{l|}{$\hat{x}^\mathrm{Weibull}_\mathrm{BLUE}(y_{1:N},\beta,\alpha) = \frac{1}{N}\sum y_k -\beta \Gamma(1+\frac{1}{\alpha})$}                      & \multicolumn{2}{c}{$\beta^2 N^{-1}\left[\Gamma(1+\frac{2}{\alpha})-\left(\Gamma(1+\frac{1}{\alpha})\right)^2\right]$}               \\[2mm]
			& \multicolumn{2}{l|}{$\hat{x}^{\mathrm{Weibull}}(y_{1:N},\beta,\alpha) =  y_{(1)} - \beta N^{-\frac{1}{\alpha}}\Gamma(1+\frac{1}{\alpha})$} & \multicolumn{2}{c}{$\beta^2 N^{-\frac{2}{\alpha}}\left[\Gamma(1+\frac{2}{\alpha})-\left(\Gamma(1+\frac{1}{\alpha})\right)^2\right]$} \\[2mm] \hline
			\multirow{2}{*}{\vspace{-3mm}$\mathrm{Pareto}(\beta,\alpha)$}             & \multicolumn{2}{l|}{$\hat{x}^\mathrm{Pareto}_\mathrm{BLUE}(y_{1:N},\beta,\alpha) = \frac{1}{N}\sum y_k -\frac{\alpha\beta}{\alpha-1}$, $\alpha>1$}              & \multicolumn{2}{c}{$\frac{\alpha\beta^2}{N(\alpha -1)^2(\alpha -2)}$, $\alpha>2$}                                                     \\[2mm]
			& \multicolumn{2}{l|}{$\hat{x}^\mathrm{Pareto}(y_{1:N},\beta,\alpha) = y_{(1)} - \frac{N\alpha\beta}{N\alpha -1}$, $N\alpha>1$}         & \multicolumn{2}{c}{$\frac{N\alpha\beta^2}{(N\alpha -1)^2(N\alpha -2)}$}                                                             
			\\[2mm] \hline
			\multirow{2}{*}{\vspace{-4mm}$\alpha\mathcal{N}(0,\sigma^2) + (1-\alpha)\mathcal{U}[0,\beta]$}             & \multicolumn{2}{l|}{$\hat{x}_\mathrm{BLUE}^{\mathcal{U},\mathcal{N}}(y_{1:N},\alpha,\beta) = \frac{1}{N}\sum_{k=1}^{N} y_k - \frac{\beta(1-\alpha)}{2}$}                                          &   \multicolumn{2}{c}{$\frac{\beta^2\left(1+\left(2-3\alpha\right)\alpha\right)+12\alpha\sigma^2}{12N}$}                                                                                                 \\[2mm]&
			\multicolumn{2}{l|}{$\hat{x}^{\mathcal{U},\mathcal{N}}(y_{1:N},\alpha,\beta) =y_{(\lfloor\frac{N\alpha}{2}\rfloor+1)}$}
			& \multicolumn{2}{c}{Unknown} \\[3mm]  \hline
	\end{tabular}}
	\label{tbl:estimators}
\end{table}

%%%%%%%%%%%%%%%%%%%%%%%%%%%%%%%%%%%%
%---------------------------SECTION----------------------------
%%%%%%%%%%%%%%%%%%%%%%%%%%%%%%%%%%%%
\section{Performance Evaluation} \label{sec:performance_evaluation}
The estimators (both unbiased and the ones without bias compensation) derived in sections~\ref{sec:distributioons_in_the_exponential_family}--\ref{sec:other_distributions}  for different noise distributions together with their MSE are summarized in Tables~\ref{tbl:estimators} and~\ref{tbl:estimators_biased}. The biased minimum order statistics based estimators and their MSE are also  The estimators derived for each noise distribution are compared against each other as a function of the sample size $N\in[2,\ldots,2000]$. Additionally, in order to verify the analytical derivations of the estimator variances, they are compared against the numerical variances obtained from $M=5000$ Monte Carlo runs. 

\subsection{Simulation Setup}
\begin{figure}[t]
	\centering
	\begin{minipage}[t]{0.47\linewidth}
		
		\includegraphics[width=1.14\linewidth]{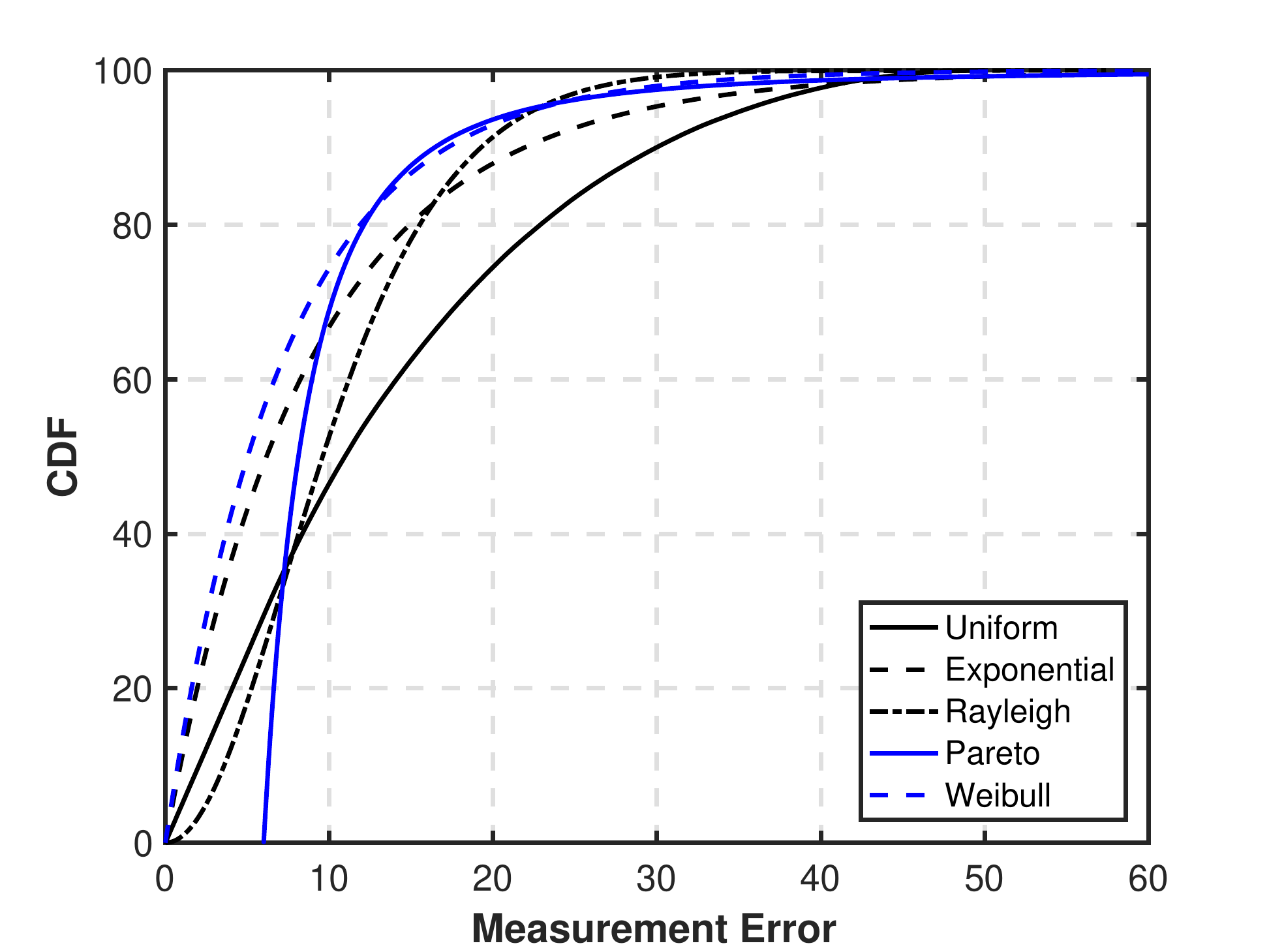}
		\caption{Empirical CDF of measurement erros computed from noise realizations used in the simulations.}
		\label{fig:cdf_error}
	\end{minipage}%
	\hfill%
	\begin{minipage}[t]{0.47\linewidth}
		\includegraphics[width=\linewidth]{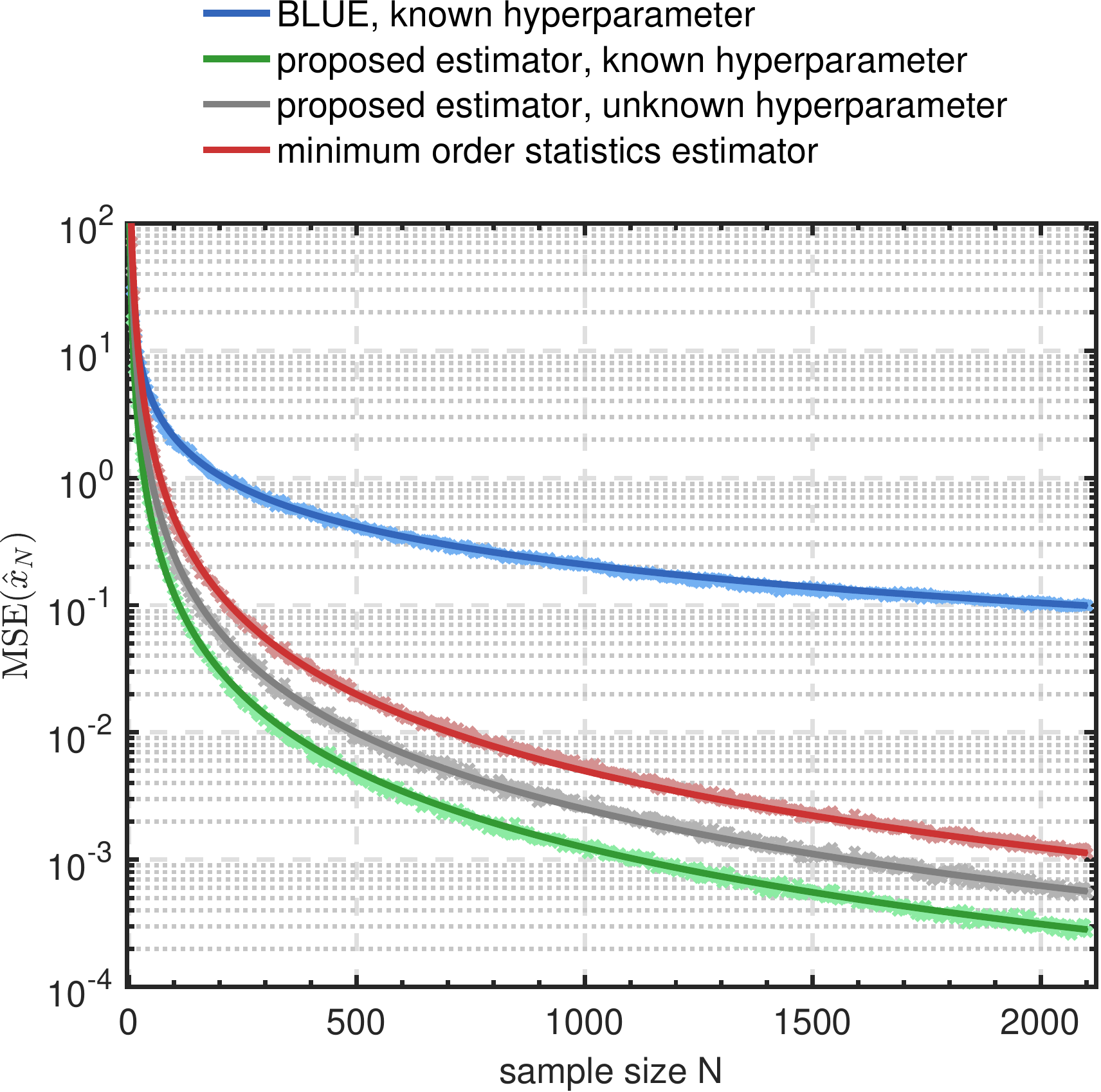}
		\caption{Analytical (marked with solid lines) and numerical (marked with crosses) MSE for uniform noise distribution as a function of the sample size $N$.}		
		\label{fig:uniform_blue_mvu}
	\end{minipage} 
\end{figure}

For each sample size, $N$ noisy measurements of the unknown parameter $x$ are generated. The hyper parameters of the noise distributions are randomly selected in each repetition. In order to have a fair comparison, the hyper parameters are randomly drawn such that the error densities are mostly in the same range for all scenarios. The noise realizations are generated from the six considered distributions with the following hyper parameters
\begin{itemize}
	\item Uniform noise: $\beta\sim\mathcal{U}[6,50]$
	\item Exponential noise: $\beta\sim\mathcal{U}[5,14]$
	\item Rayleigh noise: $\beta\sim\mathcal{U}[5,12]$
	\item Weibull noise:$\beta=1$, $\alpha\sim\mathcal{U}[5,10]$
	\item Pareto noise: $\beta=6$, $\alpha\sim\mathcal{U}[2.1,2.5]$
	\item Mixture noise: $\sigma\sim\mathcal{U}[1,9]$, $\beta\sim\mathcal{U}[1,50]$
\end{itemize} 
The empirical CDF of the error values used in the simulations are presented in Figure~\ref{fig:cdf_error}. The support of the noise values, as can be read from the figure, is $\bm{e}_m\in[0,60]$\, unit.

Let $\hat{x}_N^{(m)}$ denote the estimated value of the unknown parameter $x$  in the $m$:th repetition obtained from a sample of size $N$. For each noise distribution, the estimators' performances are evaluated in terms of the obtained MSEs. The theoretical MSE of each estimator, as defined in Table~\ref{tbl:estimators} and Table~\ref{tbl:estimators_biased}, is compared against the numerical MSE obtained in simulations. 

We let $\E[\hat{x}_N] = \frac{1}{M}\sum_{m=1}^{M}\hat{x}_N^{(m)}$ and define
\begin{subequations}
	\begin{align}
	\hat{b}_N &= \E[\hat{x}_N] - x\\
	\hat{\sigma}^2_N &= \frac{1}{M}\sum_{m=1}^{M}(\hat{x}_N^{(m)}-\E[\hat{x}_N])^2.
	\end{align}
	
	The numerical MSE for each sample size $N$ is then computed by
	\begin{align}
	\hat{\mathrm{MSE}}(\hat{x}_N) = \hat{\sigma}^2_N + \hat{b}_N^2.
	\end{align}
\end{subequations}
%
%\begin{figure}[!t] 
%	\centering
%	\includegraphics[width=0.5\textwidth,height=65mm]{cdf_error}
%	\caption{Empirical CDF of measurement erros computed from noise realizations used in the simulations.}		
%	\label{fig:cdf_error}
%\end{figure}
%
%
%
%
%\begin{figure}[h] 
%	\centering
%	\includegraphics[width=0.5\columnwidth]{estimationMse_unif}
%	\caption{Analytical (marked with solid lines) and numerical (marked with crosses) MSE for uniform noise distribution as a function of the sample size $N$.}		
%	\label{fig:uniform_blue_mvu}
%\end{figure}
%
%

\subsection{Simulation Results}
Figure~\ref{fig:uniform_blue_mvu} presents the performance of the four estimators when the noise is uniformly distributed. The solid lines correspond to the theoretical MSEs and the crosses are the numerical MSEs obtained from $M=5000$ repetitions. Both MVU estimators, with and without any knowledge of the hyper parameters of the underlying noise, result in noticeably less MSE compared to the BLUE estimator. The minimum order statistics estimator also outperforms BLUE when measurements are corrupted with additive, uniformly distributed, noise. It can be further observed that if the hyper parameter $\beta$ is unknown, the MSE of the proposed estimator is negligibly larger than the case with known $\beta$. 

\begin{figure}[!ht]
	\centering
	\begin{minipage}[t]{1\linewidth}
	\centering
\subfloat[Numerical and analytical MSE.]{\includegraphics[width=0.45\columnwidth]{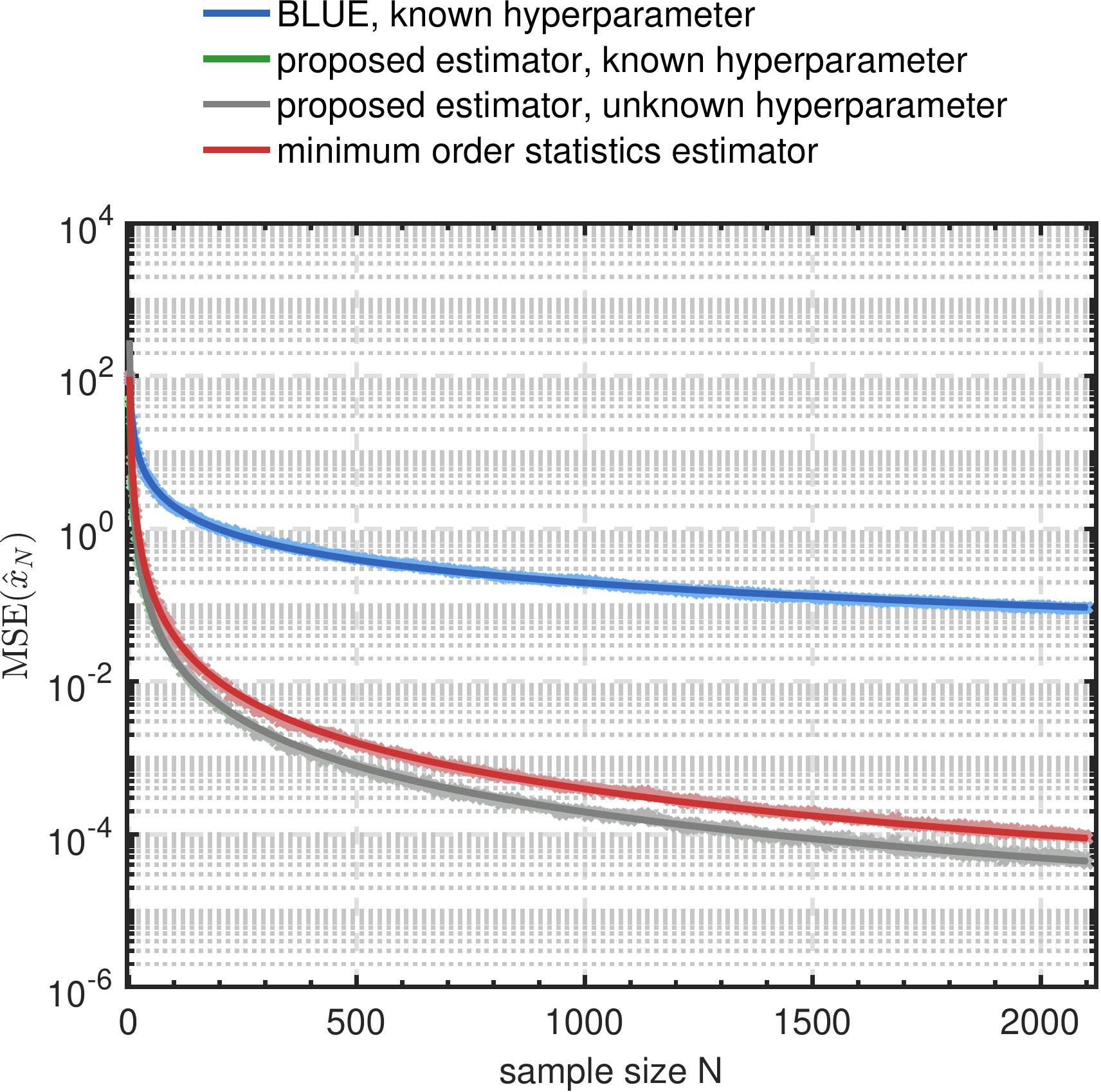}%
	\label{fig:exponential_blue_mvu_all}}
\hfil
\subfloat[Numerical and analytical MSE for $N\leq20$.]{\includegraphics[width=0.45\columnwidth]{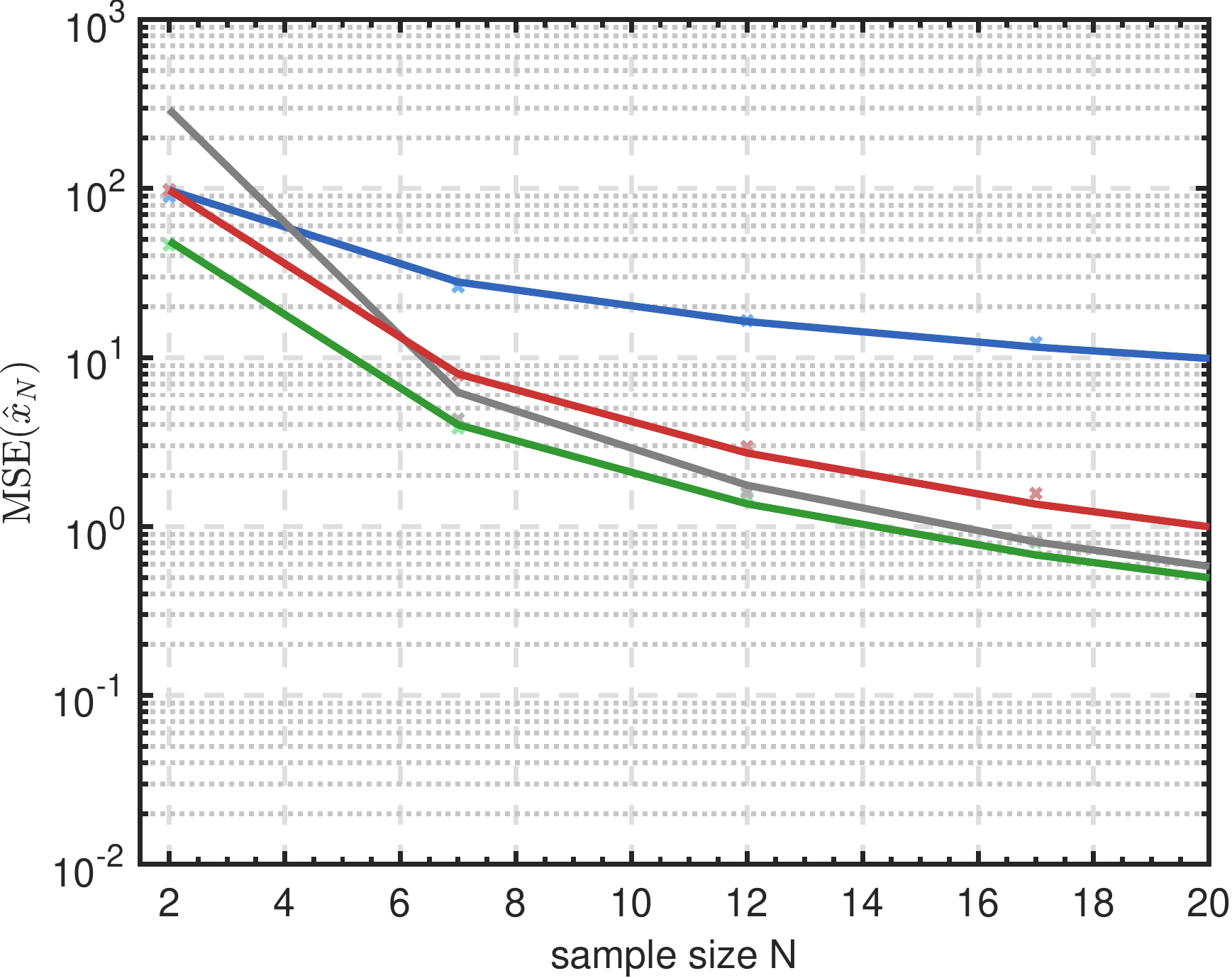}%
	\label{fig:exponential_blue_mvu_zoom}}
\caption{Analytical (marked with solid lines) and numerical (marked with crosses) MSE for exponential noise distribution as a function of the sample size $N$.  The blue and the proposed estimator with known hyperparameters have equal variances, hence the blue line is invisible in these plots. Since the MVU estimators have similar results for large sample sizes, the MSE of the four estimators for smaller sample sizes are presented separately.}
\label{fig:exponential_blue_mvu}
	\end{minipage}%
	\\
	\vspace{5mm}
	\begin{minipage}[t]{1\linewidth}
	\centering
\subfloat[Numerical and analytical MSE.]{\includegraphics[width=0.45\columnwidth]{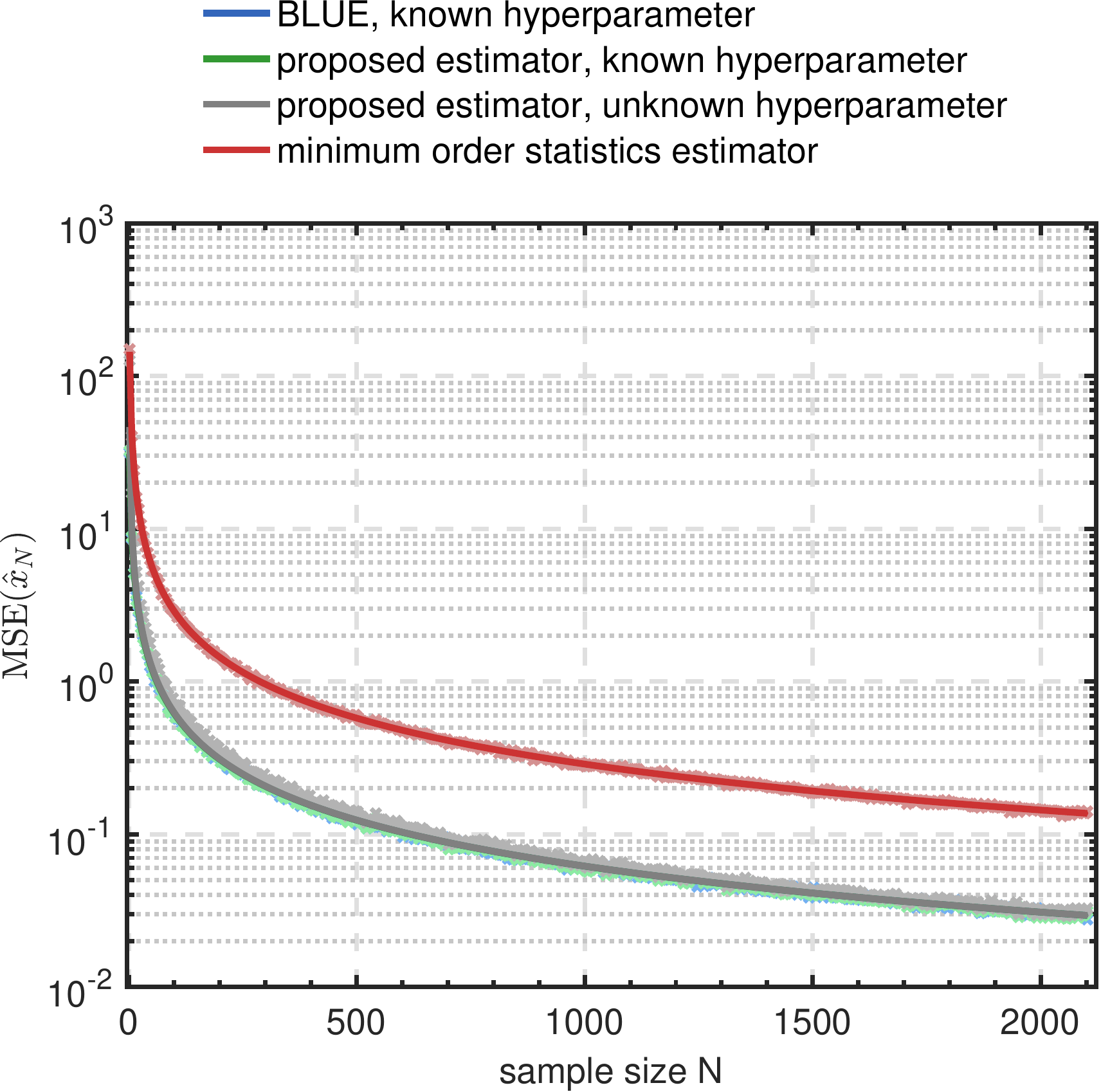}%
	\label{fig:rayleigh_blue_order_all}}
\hfil
\subfloat[Numerical and analytical MSE for $N\leq20$.]{\includegraphics[width=0.45\columnwidth]{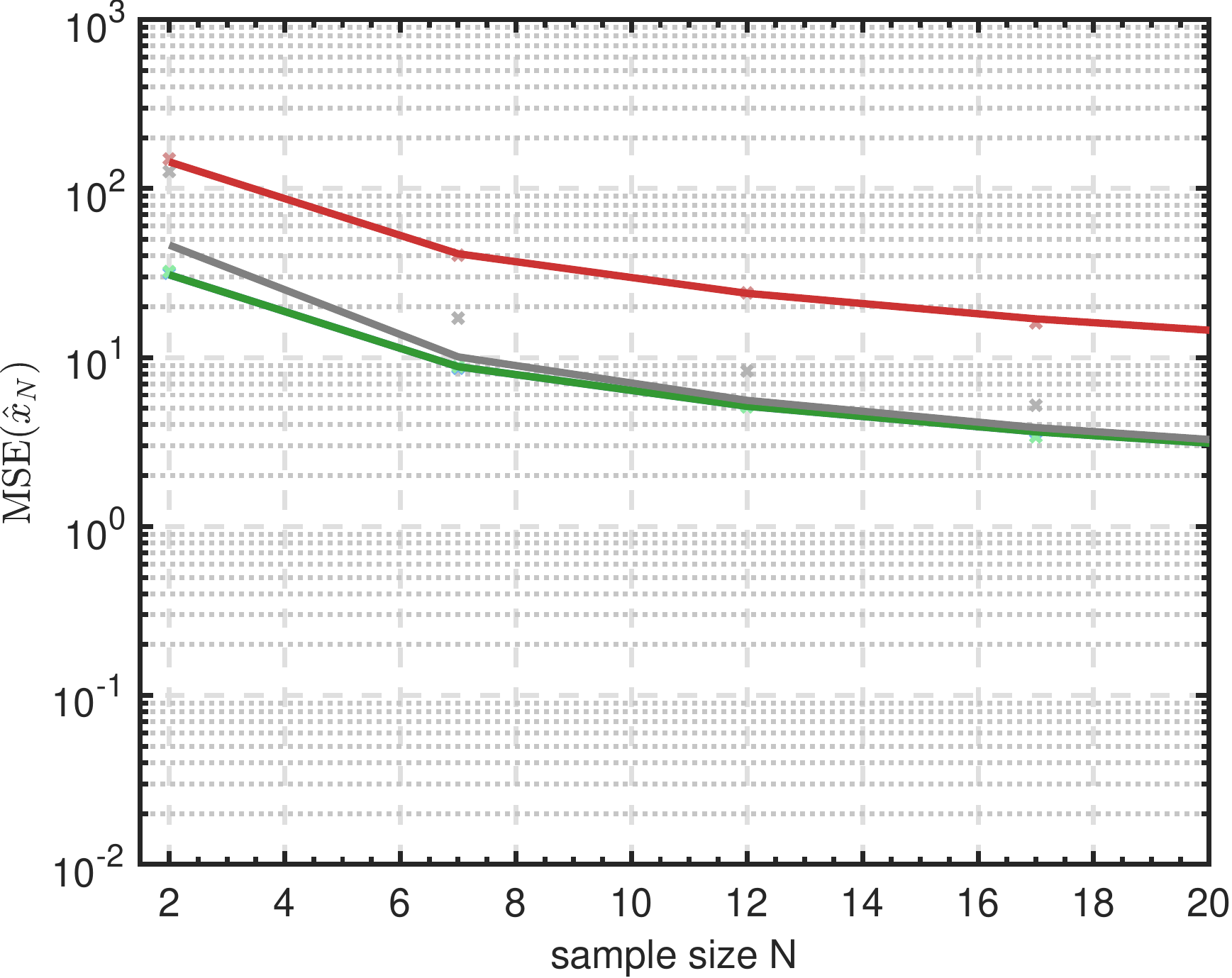}%
	\label{fig:rayleigh_blue_order_zoom}}
\caption{Analytical (marked with solid lines) and numerical (marked with crosses) MSE for Rayleigh noise distribution as a function of the sample size $N$.  Since the BLUE and the proposed estimators have similar results for large sample sizes, the MSE of the four estimators for smaller sample sizes are presented separately.}
\label{fig:rayleigh_blue_order}
	\end{minipage} 
\end{figure}
%
%
%\begin{figure}[h]
%	\centering
%	\subfloat[Numerical and analytical MSE.]{\includegraphics[width=0.5\columnwidth]{estimationMse_exponential}%
%		\label{fig:exponential_blue_mvu_all}}
%	\hfil
%	\subfloat[Numerical and analytical MSE for $N\leq20$.]{\includegraphics[width=0.5\columnwidth]{estimationMse_exponential_zoom}%
%		\label{fig:exponential_blue_mvu_zoom}}
%	\caption{Analytical (marked with solid lines) and numerical (marked with crosses) MSE for exponential noise distribution as a function of the sample size $N$.  The blue and the proposed estimator with known hyperparameters have equal variances, hence the blue line is invisible in these plots. Since the MVU estimators have similar results for large sample sizes, the MSE of the four estimators for smaller sample sizes are presented separately.}
%	\label{fig:exponential_blue_mvu}
%\end{figure}
%
%
%
%
%\begin{figure}[hb]
%	\centering
%	\subfloat[Numerical and analytical MSE.]{\includegraphics[width=0.5\columnwidth]{estimationMse_rayleigh}%
%		\label{fig:rayleigh_blue_order_all}}
%	\hfil
%	\subfloat[Numerical and analytical MSE for $N\leq20$.]{\includegraphics[width=0.5\columnwidth]{estimationMse_rayleigh_zoom}%
%		\label{fig:rayleigh_blue_order_zoom}}
%	\caption{Analytical (marked with solid lines) and numerical (marked with crosses) MSE for Rayleigh noise distribution as a function of the sample size $N$.  Since the BLUE and the proposed estimators have similar results for large sample sizes, the MSE of the four estimators for smaller sample sizes are presented separately.}
%	\label{fig:rayleigh_blue_order}
%\end{figure}
%
%
For the exponential noise distribution, as shown in Figure~\ref{fig:exponential_blue_mvu_all}, there is still a non-negligible difference between BLUE and the other three estimators in terms of estimators' MSE. However, the two MVU estimators, specially for large values of $N$, behave similarly. In order to verify their performance for smaller sample sizes, Figure~\ref{fig:exponential_blue_mvu_zoom}  illustrates the  variances of all estimators for $N\leq 20$. At the beginning, $N\in[2,4]$ the estimator with unknown hyperparameter has the largest MSE. However, for larger sample sizes, the two MVU estimators are almost equal and both have less MSE than the BLUE estimator. As in case of uniformly distributed measurement noise, the minimum order statistics estimator outperforms BLUE specially for large sample sizes.

In case of Rayleigh noise distribution, as given in Table~\ref{tbl:estimators}, the minimum order statistics estimator has the largest MSE while the BLUE and the proposed unbiased estimator with known hyper parameter, result in similar estimation variance. This can be verified also in the simulation results presented in Figure~\ref{fig:rayleigh_blue_order_all}. For large sample sizes, $N>20$, these two estimators and the proposed estimator with unknown hyperparameter  have similar values. However, for the smaller sample sizes,  as illustrated in Figure~\ref{fig:rayleigh_blue_order_zoom}, the BLUE (and order statistic with known hyper parameter) estimator has smaller variance compared to the case with unknown hyper parameter. The minimum order statistics estimator results in larger MSE compared to the other three estimators in case of Rayleigh noise distribution.

\begin{figure}[t]
	\centering
	\subfloat[Weibull noise distribution.]{\includegraphics[width=0.45\columnwidth]{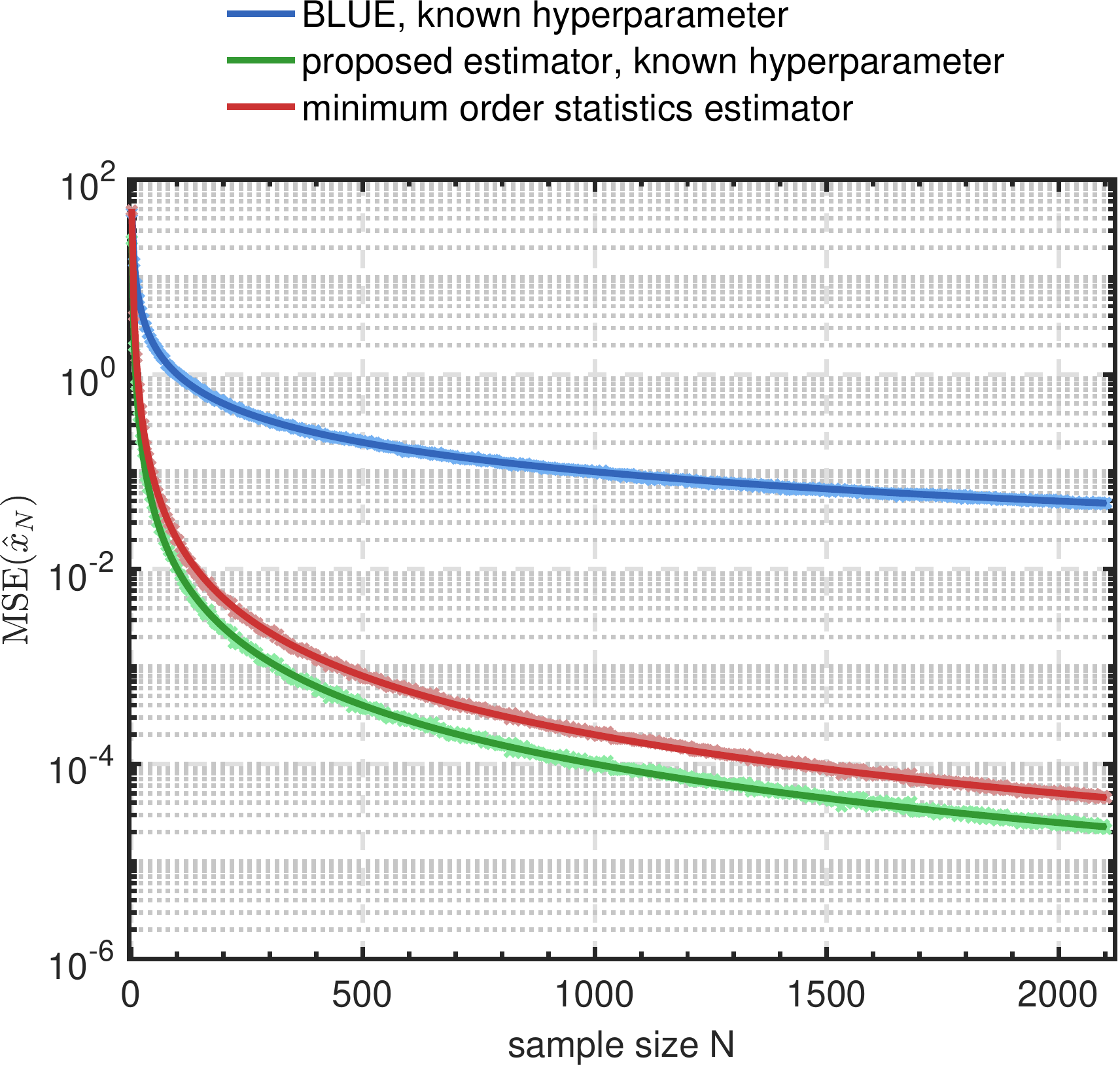}%
		\label{fig:weibull_blue_order}}
	\hfil
	\subfloat[Pareto noise distribution.]{\includegraphics[width=0.45\columnwidth]{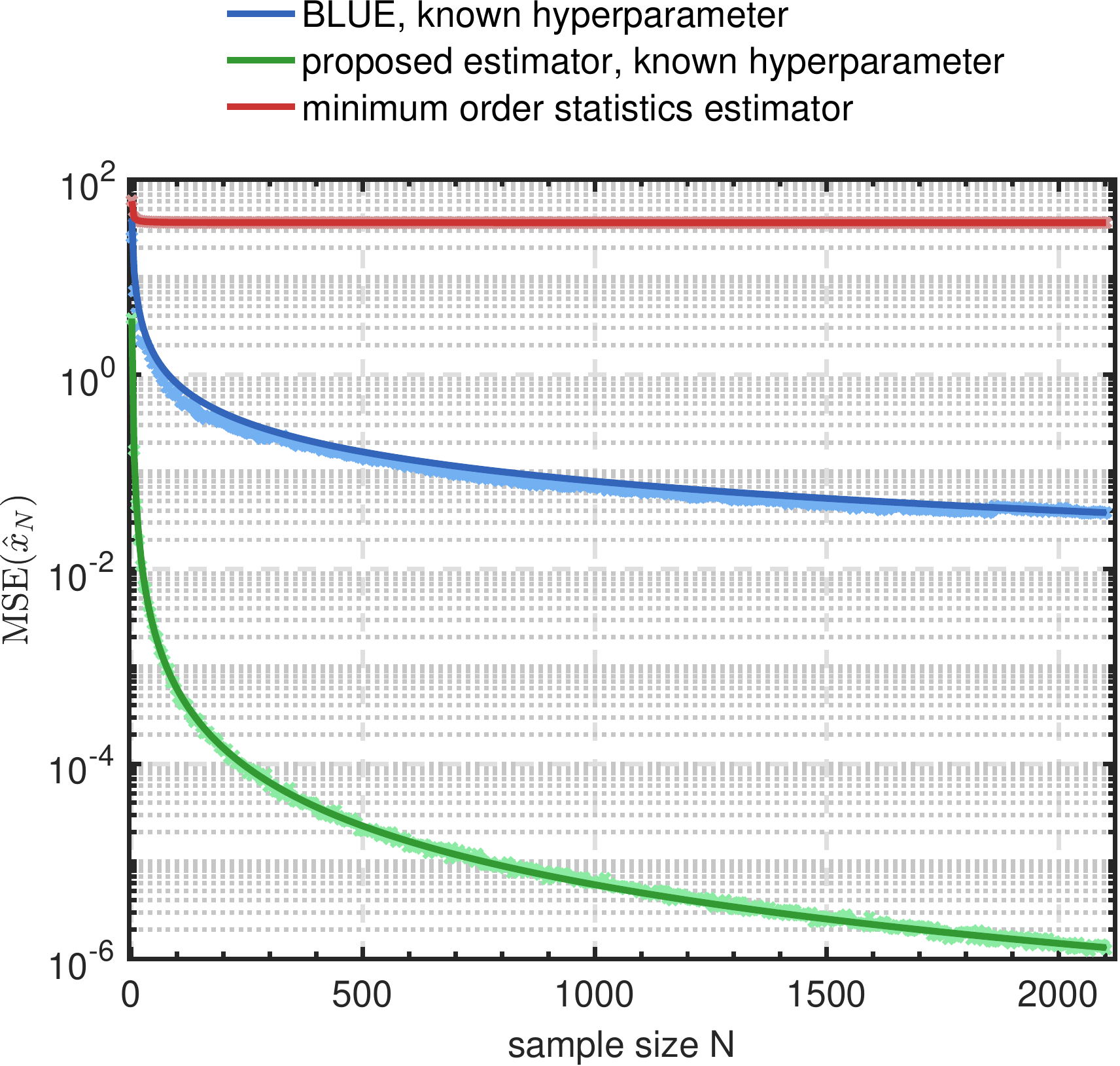}%
		\label{fig:pareto_blue_order}}
	\caption{Analytical (marked with solid lines) and numerical (marked with crosses) MSE for Weibull and Pareto noise distributions as a function of the sample size $N$.}
	\label{fig:pareto_weibull_blue_order}
\end{figure}

As Table~\ref{tbl:estimators} suggests, for Pareto and Weibull noise distributions, we only derived BLUE and an unbiased order statistics based estimators when the two hyperparameters of the distributions are known. For both noise distributions, the MSE of the two unbiased estimators as well as the MSE of the minimum order statistics estimator are compared and the results are presented in Figure.~\ref{fig:pareto_weibull_blue_order}. In both cases, the proposed estimators outperform the BLUE in terms of variance. The minimum order statistics estimator results in a lower MSE than the BLUE for Weibull noise distributions. However, in case of Pareto noise, the BLUE has a better performance compared to the minimum order statistics estimator.
\begin{figure}[b]
	\centering
	\subfloat[Histogram of noise realizations and the fitted PDF.]{\includegraphics[width=0.45\columnwidth]{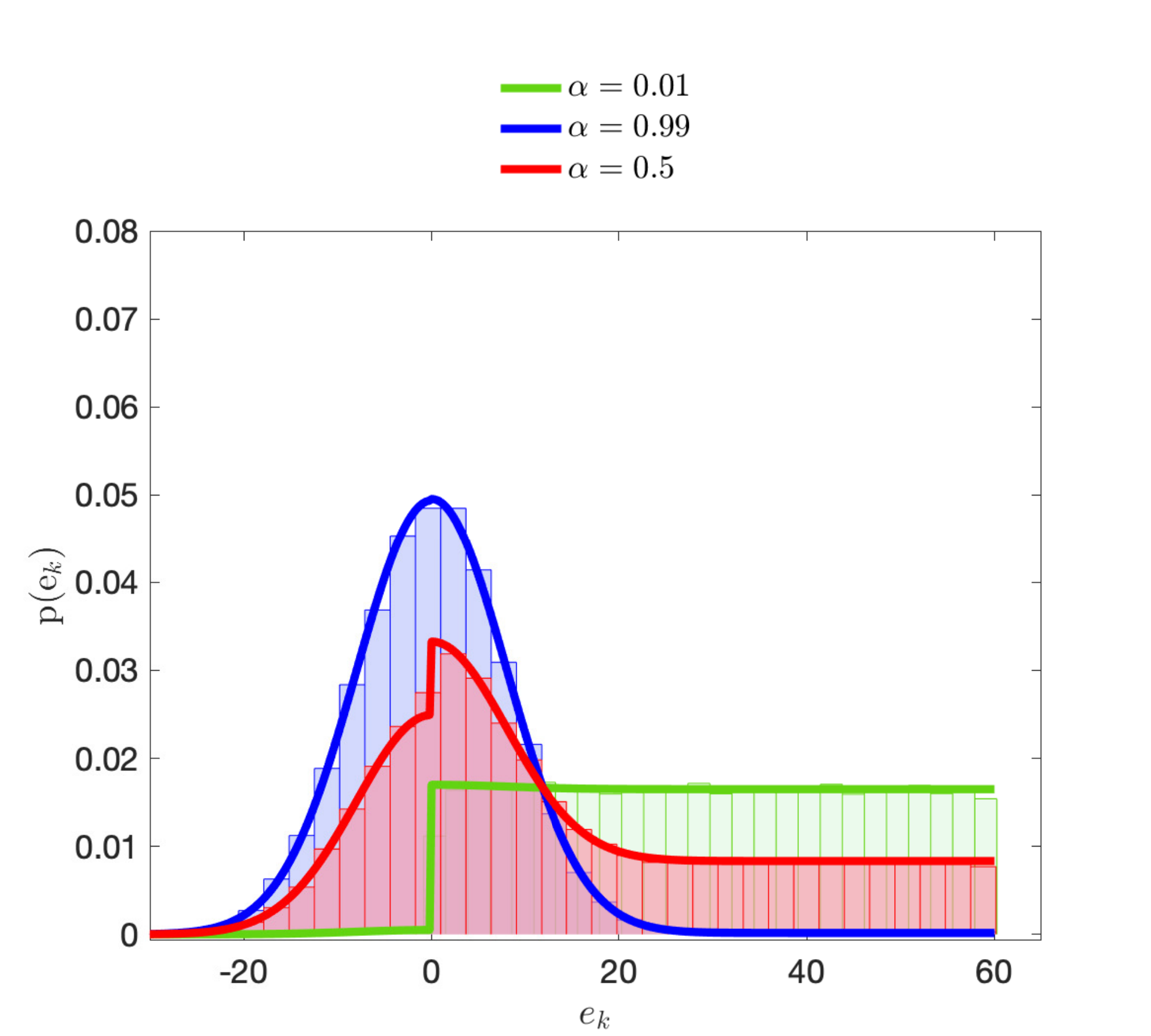}%
		\label{fig:mixture_pdf}}
	\hfil
	\subfloat[Empirical CDF.]{\includegraphics[width=0.45\columnwidth]{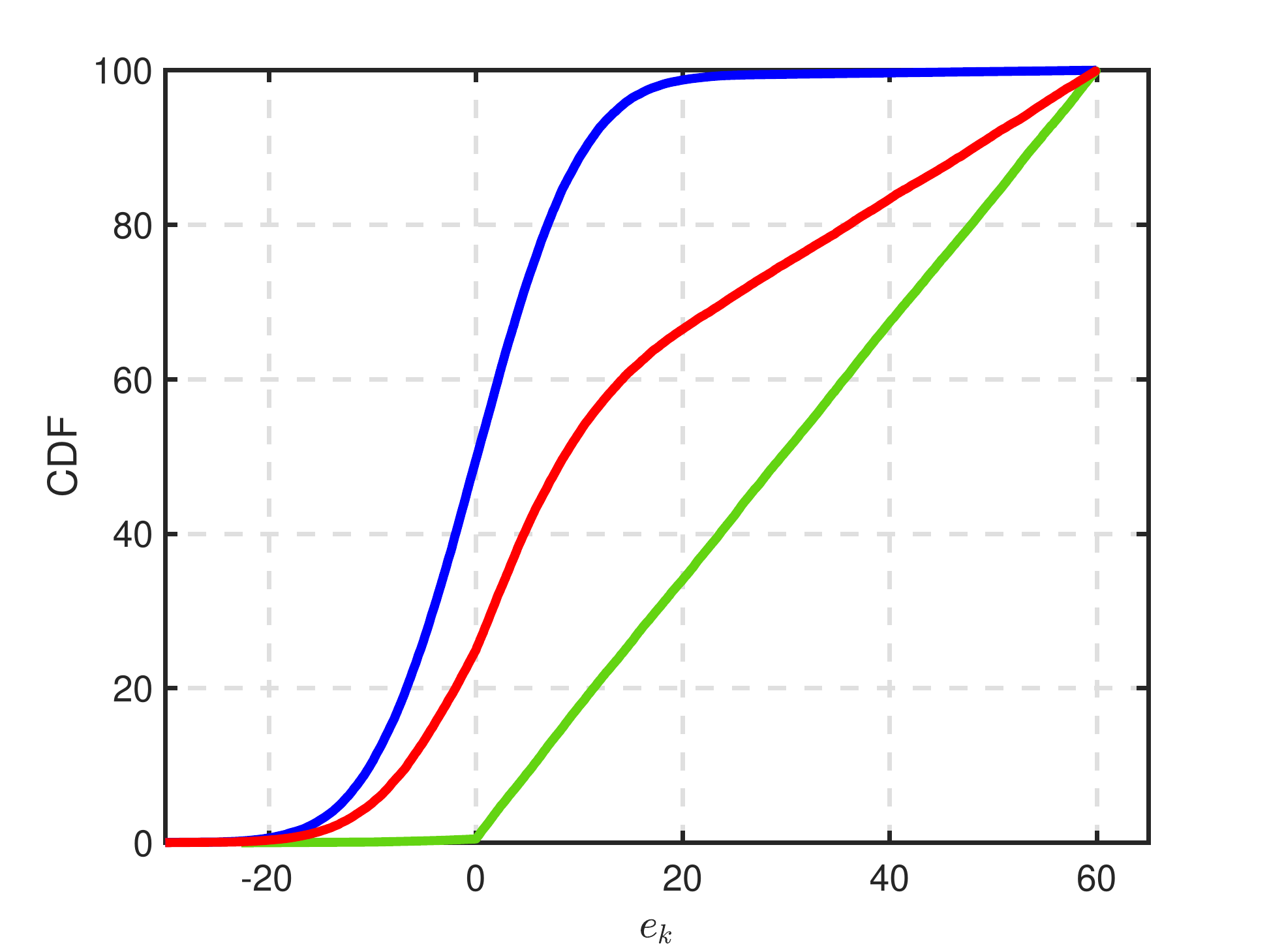}%
		\label{fig:mixture_cdf}}
	\caption{Noise realizations, PDF, and empirical CDF of the mixture noise distribution $e_k\sim \alpha\mathcal{N}(0,8^2) + (1-\alpha)\mathcal{U}(0,60)$ for three different values of $\alpha$.}
	\label{fig:mixture_noise}
\end{figure}

In case of mixture noise distribution, we consider three different scenarios based on the mixing probabilities; two extreme cases with dominant contribution from uniform noise, $\alpha=0.01$, and dominant contribution from normal noise, $\alpha=0.99$, and the case with $\alpha=0.5$. Fig.~\ref{fig:mixture_pdf} illustrates the histogram of the noise realizations of the considered mixture noise distributions $e_k\sim \alpha\mathcal{N}(0,8^2) + (1-\alpha)\mathcal{U}(0,60)$ and the fitted densities.  The empirical CDFs of the errors for the three cases are presented in Figure~\ref{fig:mixture_cdf}. 

%%
%%
%\begin{figure}[t] 
%	\centering
%	\includegraphics[width=0.5\textwidth]{cdf_mixture}
%	\caption{.}		
%	\label{fig:mixture_cdf}
%\end{figure}
%%
%%

\begin{figure}[]
	\centering
	\subfloat[$\alpha=0.01$.]{\includegraphics[width=0.33\columnwidth]{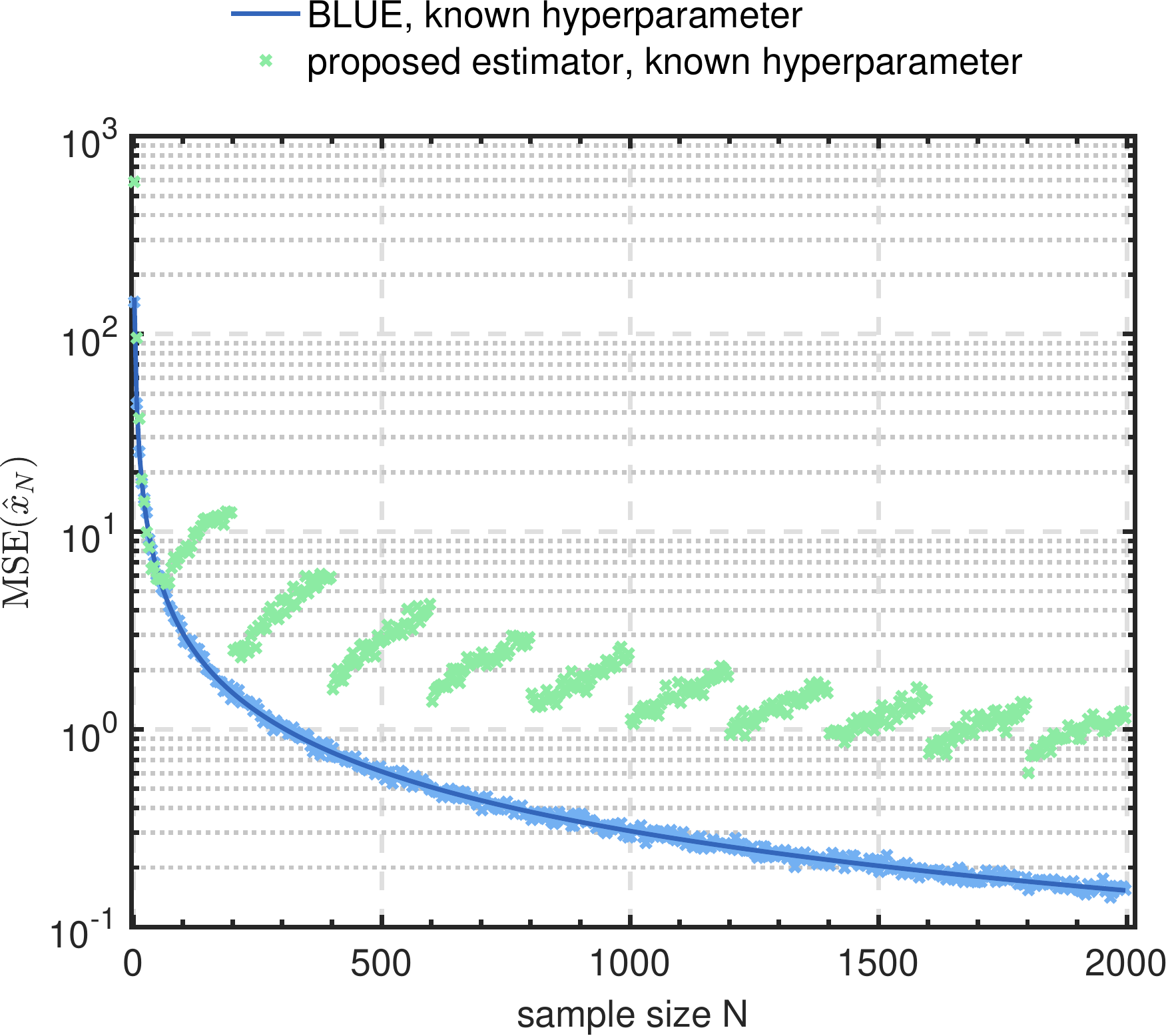}%
		\label{fig:estimationMse_mixture01}}
	\hfil
	\subfloat[$\alpha=0.99$.]{\includegraphics[width=0.33\columnwidth]{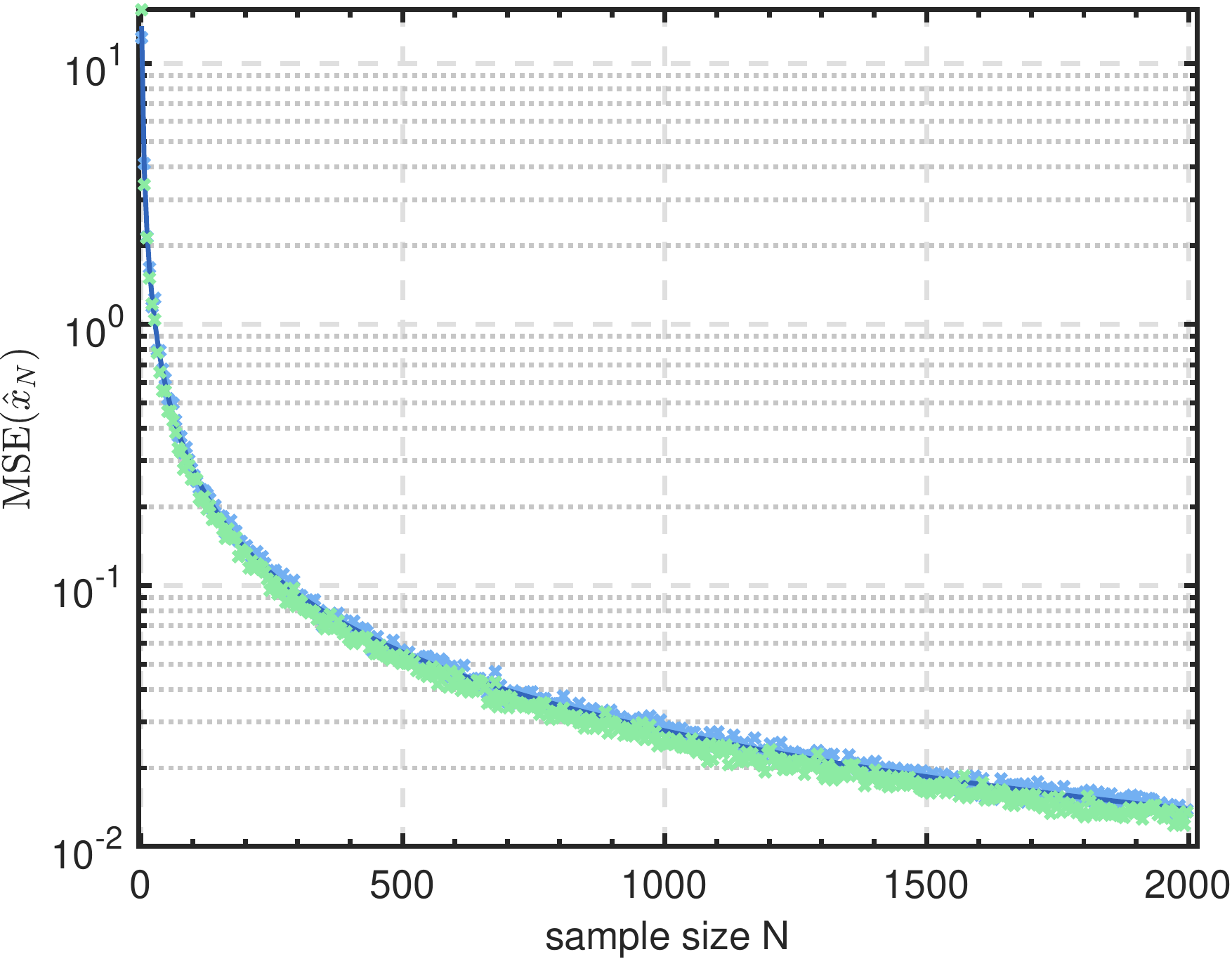}%
		\label{fig:estimationMse_mixture99}}
		\hfil
	\subfloat[$\alpha=0.5$.]{\includegraphics[width=0.33\columnwidth]{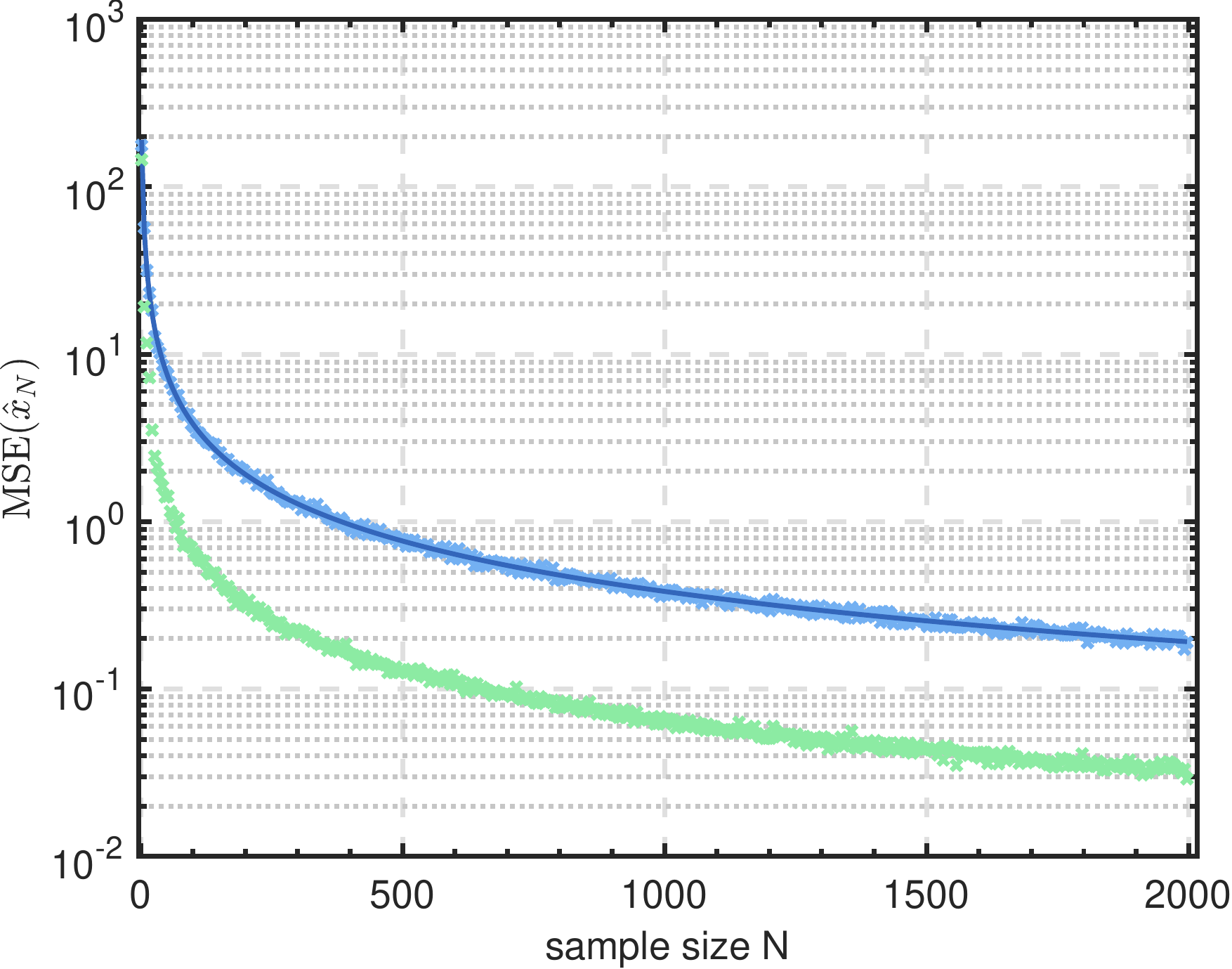}%
		\label{fig:estimationMse_mixture5}}
	\caption{Analytical (marked with solid lines) and numerical (marked with crosses) MSE for three different values of mixing probability $\alpha$, when $e_k\sim \alpha\mathcal{N}(0,8^2) + (1-\alpha)\mathcal{U}(0,60)$.}
	\label{fig:mixture_order}
\end{figure}

In order to estimate the unknown parameter $x$, in each Monte Carlo run, we sort the measurements and then find the $(\lfloor\frac{N\alpha}{2}\rfloor+1)$:th component. Figure~\ref{fig:mixture_order} presents the estimation MSE for the three different scenarios with different mixing probabilities. As the results indicate, when the main contribution of the noise is from uniform distribution, $\alpha=0.01$, BLUE outperforms the proposed estimator.  In this case, a periodic behavior for the MSE can be observed. The jumps in the MSEe occur exactly at pints where $\lfloor\frac{N\alpha}{2}\rfloor+1$ switches from the $k$:th measurement to the $k+1$:th measurement. For instance, for $N\in[1,199]$, $\lfloor\frac{N\alpha}{2}\rfloor=0$, hence $\hat{x}=y_{(1)}$. However, at $N=200$, $\lfloor\frac{N\alpha}{2}\rfloor=1$, resulting in $\hat{x}=y_{(2)}$. 

The proposed estimator and the BLUE result in similar estimation MSE for $\alpha=0.99$, as shown in Figure~\ref{fig:estimationMse_mixture99}, in which the normal component is the dominant source of error. However, the most interesting results are obtained when both distributions have equal contributions in the measurement noise, {\em i.e} $\alpha=0.5$. In this case, as Figure~\ref{fig:estimationMse_mixture5} suggests, the proposed estimator outperforms the BLUE.
\section{Conclusions}\label{sec:conclusions}
In this work, the location estimation problem was studied in which an unknown parameter was estimated from observations under additive noise. Multiple noise distributions were considered and, in some cases, MVU estimators were proposed. In other cases an unbiased estimator based on minimum order statistic was derived. Furthermore, if applicable, MVU and minimum order statistic estimators without any knowledge of the hyper parameters of the underlying noise distributions were provided. The results of all the estimators were compared with BLUE in terms of variance for various measurement sample sizes. The results indicate better performance of the proposed estimators compared to BLUE. Additionally, the location estimation problem under mixture of normal and uniform noise distribution was studied and the numerical MSE of the proposed estimator were evaluated. The simulation results indicate that for the extreme cases where either of the two components, Gaussian or uniform, are dominant, the proposed estimator cannot beat the BLUE. However, when the mixing probability is not in the extreme region, {\em e.g} larger than $1$ percent, the proposed estimator has a noticeably less MSE compared to the BLUE. 

%\bibliographystyle{unsrt}
%\bibliography{references}
%\bibliographystyle{IEEEtran}
\bibliographystyle{unsrtnat}
\bibliography{Exploring-positive-noise-in-estimation-theory}

\end{document}